\documentclass[11pt, letterpaper]{article}

\usepackage{graphicx}
\usepackage{amsmath}
\usepackage{amssymb}
\usepackage{mathtools}
\usepackage{setspace}
\usepackage{enumitem} 
\usepackage{epstopdf}
\usepackage{color}
\usepackage[letterpaper,left=1in,right=1in,top=1in,bottom=1in]{geometry}
\usepackage{algorithm}
\usepackage{algpseudocode}
\usepackage{multirow}
\usepackage{longtable}
\usepackage{rotating}
\usepackage{mathrsfs}
\usepackage{subfig}
\usepackage{bm}
\usepackage{float}
\usepackage{graphicx}
\usepackage{caption}
\usepackage{amsthm}
\usepackage{xcolor}

\newtheorem{remark}{Remark}

\graphicspath{{Figures_new}}

\title{\Large \bf LSTM-based model predictive control with discrete inputs for irrigation scheduling}

\author{
\centerline{\normalsize Bernard T. Agyeman$^{a}$, Soumya R. Sahoo$^{a}$, Jinfeng Liu$^{a}$, Sirish L. Shah$^{a}$
}
\vspace{5mm}\\
\centerline{\small $^{a}$Department of Chemical \& Materials Engineering, University of Alberta,}\\
\centerline{\small Edmonton, AB T6G 1H9, Canada.}}
\allowdisplaybreaks

\begin{document}

\date{}
\maketitle

\setstretch{1.5}

\begin{abstract}
The development of well-devised irrigation scheduling methods is desirable from the perspectives of plant quality and water conservation. In this article, a model predictive control (MPC) with discrete actuators is developed for irrigation scheduling, where a long short-term memory (LSTM) model of the soil-water-atmosphere system is used to evaluate the objective of ensuring optimal water uptake in crops while minimizing total water consumption and irrigation costs. A heuristic method involving a sigmoid function is used in this framework to enhance the computational efficiency of the scheduler. The scheduling scheme is applied to homogeneous and spatially variable fields and the results indicate that the LSTM-based MPC with discrete actuators is able to prescribe optimal or near-optimal irrigation schedules that are typical of irrigation practice. 
\end{abstract}


\section{Introduction}
According to the United Nations, agriculture accounts for about 70\% of global freshwater withdrawals, the vast majority of which are used for irrigation purposes~\cite{UN_report}. At the same time, the global water scarcity crisis is worsening, due to increased stress on freshwater resources resulting from population growth and climate change. Given the rising freshwater shortages, there is a pressing need for enhanced and precise irrigation management strategies that will enable efficient and sustainable water use while ensuring optimal plant development.

The importance of water in plants stems from its vital role in photosynthesis, plant temperature regulation, and nutrient transport. While irrigation ensures plant growth in areas where rainfall is not enough to support optimal plant yield, it is beneficial to irrigate to meet the specific water requirements of plants at the right time instant while avoiding under and over irrigation. This can be achieved by implementing well-devised irrigation control and scheduling operations on an hourly or daily basis for a planning horizon of usually up to a few hours, days or weeks~\cite{ali2001methods}. Traditionally, most control and scheduling operations in irrigation management are implemented in an open-loop fashion, where there is no direct connection between the supplied irrigation volume and the prevailing soil water status. In an open-loop structure, the water needs of crops are determined using climatic factors, soil texture, and heuristics created by farm operators. For instance, most irrigation schedulers are programmed to supply a fixed amount of water at a specific time instant. Additionally, some schedulers are designed to irrigate after a predetermined soil moisture content is attained~\cite{pardossi2011traditional}. Such open-loop implementations are imprecise and thus do not guarantee optimal plant yield and enhanced water use efficiency. Precision irrigation methods have been advocated as a means of alleviating the drawbacks that are associated with open-loop irrigation operations. In the context of systems engineering, precision irrigation can be realized by closing the irrigation decision support loop to form a closed-loop system~\cite{shah2021meeting}. Closed-loop irrigation controllers and schedulers  employ sensor feedback to precisely determine the irrigation volume and the timing of the irrigation event. Precision irrigation has been demonstrated to enhance water use efficiency, reduce irrigation costs, and enhance crop yields~\cite{navarro2015wireless,monaghan2013more,hedley2009soil,morillo2015toward}.

Irrigation scheduling seeks to provide crops with the right amount of water at appropriate times.  Among all the irrigation scheduling methods that have been recommended and developed, four main types can be clearly distinguished: (1) evapotranspiration and soil water balance-based (ET-SWB) scheduling, (2) soil water status-based (SWS) scheduling, (3) plant water status-based (PWS) scheduling, and (4) model-based irrigation scheduling~\cite{gu2020irrigation}.
In ET-SWB-based methods~\cite{sethi2006optimal,khare2007assessment,bartlett2015smartphone,perea2017multiplatform}, climatic variables, such as total absorbed radiation, relative humidity, air temperature, and wind speed, are used to estimate  crop evapotranspiration $ET_c$. Using the estimated $ET_c$ and a soil water balance equation, the soil water deficit can be inferred. Irrigation events are scheduled whenever the inferred water deficit exceeds a threshold value. 
While ET-SWB-based methods are computationally efficient, they are unable to capture crop water requirements and crop growth in a realistic manner due to inaccuracies in the estimation of $ET_c$~\cite{nguyen2017optimization}. In SWS-based methods~\cite{hedley2009method,thompson2007using,gutierrez2013automated,haule2014deployment}, measured soil water content, which is inferred from sensor measurements, is compared to a threshold in order to trigger the irrigation event. These computationally efficient methods primarily determine when the irrigation event should be performed in order to maximize crop yield. Additionally, SWS-based methods permit variable rate irrigation scheduling due their ability to capture the temporal and spatial variability of soil moisture. However, these methods are unable to capture the water needs of crops correctly due to inaccurate measurements from the sensors~\cite{evett1996canopy}. SWS-based methods are also relatively less practical because of the need to deploy and maintain a large number of sensors in the field. In PWS-based methods~\cite{dejonge2015comparison,padilla2016scheduling,osroosh2015automatic,bartlett2015smartphone}, stress variables that characterize the need for irrigation, such as crop water stress index, the stem diameter variation index, and the crop canopy temperature are measured or visually inferred. While PWS-based methods are easy to implement, inaccuracies in the measurement of these  stress indices often lead to a false triggering of the irrigation scheduler.

To obtain a more precise, dependable, and robust scheduler, models that capture the dynamics of the soil-crop-atmosphere system (agro-hydrological models) have been used to  determine irrigation schedules. In model-based methods, a calibrated process model, which typically employs feedback from soil and climatic sensors, is used to schedule irrigation~\cite{park2009receding}. For example Lopes et al.~\cite{lopes2016optimal} transformed the scheduling problem into an optimal control problem with the objective of minimizing the total volume of irrigation during the entire growing season. This optimal control problem relied on a soil water balance model. Nguyen et al.~\cite{nguyen2017optimization} developed an ant colony method to search for optimal irrigation schedules. This ant colony method was developed with the Root Zone Water Quality Model (RZWQM2). Park et al.~\cite{park2009receding} used the one-dimensional (1D) Richards equation to develop a model predictive control (MPC) system that provided irrigation schedules for a center pivot irrigation system. Delgoda et al.~\cite{delgoda2016irrigation} combined an MPC system with the AquaCrop model to determine the irrigation volume. However, the use of  mechanistic models, also known as first principle models, in ~\cite{lopes2016optimal,nguyen2017optimization,park2009receding,delgoda2016irrigation} has a number of practical drawbacks. These models are often  more difficult to handle from a numerical point of view and hence they render the resulting scheduling scheme computationally inefficient. Furthermore, mechanistic models require time-consuming calibrations during their development.

Over the past few years, many studies have examined the use of statistical or data-driven models, also known as black box models, in irrigation scheduling. Nahar et al.~\cite{nahar2019closed} identified a linear parameter varying (LPV) model for a closed-loop scheduler and controller using input and output data obtained from the 1D Richards equation. However, due to the inability of LPV models to accurately describe highly nonlinear systems, such as agro-hydrological systems, there was a significant mismatch between the predictions of the identified LPV model and that of the Richards equation. Data-driven machine learning approaches such as adaptive neuro-fuzzy inference systems (ANFIS)~\cite{karandish2016comparison} and  support vector machines (SVMs)~\cite{karandish2016comparison,deng2011soil,liu2010data} are another group of statistical models that have been used to develop model-based irrigation schedulers. While these models have demonstrated prediction performances comparable to that of the HYDRUS-2D model, they are less robust and also generalize poorly due to their inability to make predictions under water stress conditions~\cite{karandish2016comparison}.

Neural network models have been used to predict the dynamics of soil moisture due to their strong learning potential and their capacity to model nonlinear relationships between inputs and outputs. In Capraro et al.~\cite{capraro2008neural}, Tsang and Jim~\cite{tsang2016applying}, and Gu et al.~\cite{gu2021neural}, irrigation schedules were determined based on soil moisture predictions from a feedforward neural network. Feedforward neural networks are generally unable to model dynamic data due to their inability to preserve past information. Therefore, when these networks are applied to highly causal systems, such as agro-hydrological systems, they result in suboptimal predictions~\cite{brezak2012comparison}. To improve the ability of feedforward networks to model dynamic data, Pulido-Calvo and Gutiérrez-Estrada~\cite{pulido2009improved} corrected soil moisture predictions obtained from feedforward networks with a fuzzy logic whose parameters were tuned with genetic algorithms. While this approach resulted in improved soil moisture predictions, the use of improvised fuzzy logic rules limited the generalization ability of the resulting feedforward neural network. Recurrent neural networks (RNNs) are a robust type of neural network that are specifically designed to handle sequential data. Because of their internal memory, they can remember information from previous time steps, which allows them to be very accurate in modeling dynamic data. Theoretically, RNNs can be trained to solve problems that require learning long-term temporal dependencies. However, in practice, they are unable to learn such long-term dependencies due to the vanishing/exploding gradient problem. Long short-term memory (LSTM) are a special kind of RNN that are capable of learning long-term temporal dependencies and they have been successfully applied to nonlinear dynamical systems. In the context of irrigation scheduling, Adeyemi et al.~\cite{adeyemi2018dynamic} developed an LSTM model for the prediction of soil moisture content. This LSTM model was combined with the ET-SWB-based scheduling method to determine the site-specific irrigation amount on a daily basis as well as the timing of the irrigation event.

Many optimal control approaches have been applied to schedule irrigation. In the context of irrigation scheduling, optimal control methods determine the  control variables, specifically the water application depth and irrigation time, that minimize or maximize some performance measure. The formulation of an optimal control problem requires a simulation model of the process to be controlled, in this case a model of the agro-hydrological system; a statement of the physical constraints of the control and controlled variables; and a specification of the performance measure to be optimized, such as net return to the farm operator, seasonal yield, energy requirements, total irrigation amount, among others. For example in Rao et al,~\cite{rao1988irrigation} and Naadimuthu et al.~\cite{naadimuthu1999heuristic}, irrigation was scheduled using the dynamic programming method to maximize crop yield. Although the dynamic programming approach has the advantage of being simple for unconstrained linear systems with a small state dimension, it is suffers from the ``curse of dimensionality'' (i.e., the size of the search space grows exponentially with the number of state variables) which limits its practical application in irrigation scheduling.
Model predictive control (MPC) is regarded to be an effective means of implementing the dynamic programming solution~\cite{rawlings2017modelbook}. MPC relies on a sufficiently descriptive model of a system to optimize some performance measure and ensure constraint satisfaction. In an MPC scheme, at each sampling time instant, a finite horizon optimal control problem is solved in which the initial state is the current state of the system. This optimization yields a finite control sequence and the first control action is applied to the system. This is repeated at each sampling time instant with a receding prediction horizon. MPC has demonstrated remarkable success for the high-performance control of technical systems in a wide range of applications such as process systems control, robotics, automotive applications, irrigation applications, and vehicle path planning~\cite{mayne2014model}. For example, in the area of irrigation scheduling and control, McCarthy et al.~\cite{mccarthy2014simulation} used MPC to determine irrigation timing and site-specific irrigation volumes on a daily basis. Similarly, in Delgolda et al.~\cite{delgoda2016irrigation}, irrigation was scheduled using MPC to minimize the root zone soil moisture deficit and the total irrigation volume. In these studies, the MPCs were designed to maintain the root zone soil moisture at a predetermined set-point. However, these set-point MPCs can be overly stringent. It is sufficient to keep soil moisture content within a range/zone when it comes to ensuring optimal water uptake by crops. Usually, the upper bound of the desired range is a point below the field capacity and the lower bound is a point above the permanent wilting point. A model predictive control with zone objectives (zone control) is thus a natural choice for irrigation scheduling. From a theoretical point of view, the control objective of an MPC with zone control can be thought of as tracking a target set in the output space since there are no preferences between points in the target zone~\cite{ferramosca2010mpc}. MPC with zone objectives has seen applications in different areas such as diabetes treatment~\cite{gondhalekar2013periodic,gondhalekar2016periodic}, building heat control system~\cite{privara2011model}, and pressure management of a water supply network~\cite{liu2016predictive}, and soil moisture regulation~\cite{nahar2019closedzone}. In Nahar et al.~\cite{nahar2019closedzone}, closed-loop irrigation scheduling was performed using a zone MPC to maximize crop yield.

Scheduling problems are inherently combinatorial in nature because the allocation of scarce resources to competing tasks over time involves many discrete decisions~\cite{floudas2005mixed}. When irrigation is  to be scheduled on a daily basis, the determination of the irrigation time reduces to a discrete decision of whether or not the irrigation event should be performed on each day within the planning horizon. Thus, the daily irrigation scheduling problem can be transformed into an optimal control problem with both continuous- (irrigation volume or depth) and integer- (irrigation time) valued control variables. The early literature on MPC focused almost entirely on continuous-valued controls due in part to the computational difficulty imposed by discrete decision variables~\cite{mayne2000constrained,garcia1989model}. In industrial practice, discrete decisions are always excluded from the MPC control formulation and rather made at a separate layer of the control hierarchy using heuristics~\cite{rawlings2017model}. With the improvements in optimization software and computing performance, however, discrete-valued controls can be selected optimally by directly including them in the MPC design. The incorporation of discrete decisions in the MPC design does not change the structure of the MPC  and any result that holds for the standard MPC holds also for MPC with discrete-valued control variables~\cite{rawlings2017model}. MPC with discrete actuators has seen applications in different areas such as energy systems~\cite{risbeck_2018}, heat pumps~\cite{lee2019mixed} and space robots~\cite{zong2018obstacle} and theoretical studies on MPC with discrete actuators can be found in~\cite{rawlings2017model}. As in ~\cite{risbeck_2018, zong2018obstacle, rawlings2017model}, we refer to MPC with both continuous- and discrete-valued control variables as a mixed-integer MPC in the rest of this paper.

Motivated by the above, this work develops an irrigation scheduler in the framework of a mixed-integer MPC with zone control for agro-hydrological systems that utilize an LSTM model to predict the dynamics of soil moisture. Specifically, the LSTM model is initially developed based on a dataset generated from extensive open-loop simulations of a mechanistic agro-hydrological model. Subsequently, a mixed-integer MPC with zone objectives that utilizes the identified LSTM model as its prediction model is developed to determine irrigation schedules. The proposed irrigation scheduler calculates the irrigation time and the irrigation depth that ensure optimal root water uptake while minimizing irrigation costs and total water consumption. Due to the inherently complex nature of mixed-integer programs, this work further proposes a heuristic method that can be used to simplify the mixed-integer MPC in order to reduce its computation time. The main contributions of this work include: 
\begin{enumerate}
\item A method to identify an LSTM model for the prediction of soil moisture content in an agro-hydrological system for both uniform and spatially variable fields.
\item A detailed closed-loop irrigation scheduler design in the framework of a mixed-integer MPC with zone control that ensures optimal root water uptake while minimizing irrigation costs and total water consumption. This work proposes separate designs for uniform and spatially variable fields. In the context of irrigation scheduling in spatially variable fields, designs that are tailored to small- and large-scale fields are also developed.
\item A heuristic method, using a sigmoid function,  that simplifies the mixed-integer MPC in order to reduce its computation time. 
\item Extensive simulations that investigate the performance of the proposed scheduling approach. The performance of the proposed heuristic method is also demonstrated with simulations.
\end{enumerate}

Elements of the work presented in this article were reported in~\cite{agyeman2021}. Compared with~\cite{agyeman2021}, this article presents detailed explanations, comprehensive set of results,  and it introduces new set of results for irrigation scheduling in small- and large-scale spatially variable fields
\section{Preliminaries}
In this section, we first provide a description of the soil-water-atmosphere system and introduce a mechanistic equation that can be used to model the soil moisture dynamics in an agro-hydrological system. This section concludes with a theoretical background on LSTMs. 
\subsection{Agro-hydrological system}
\begin{figure}[H]
	\centering
	\centerline{\includegraphics[width=0.4\textwidth]{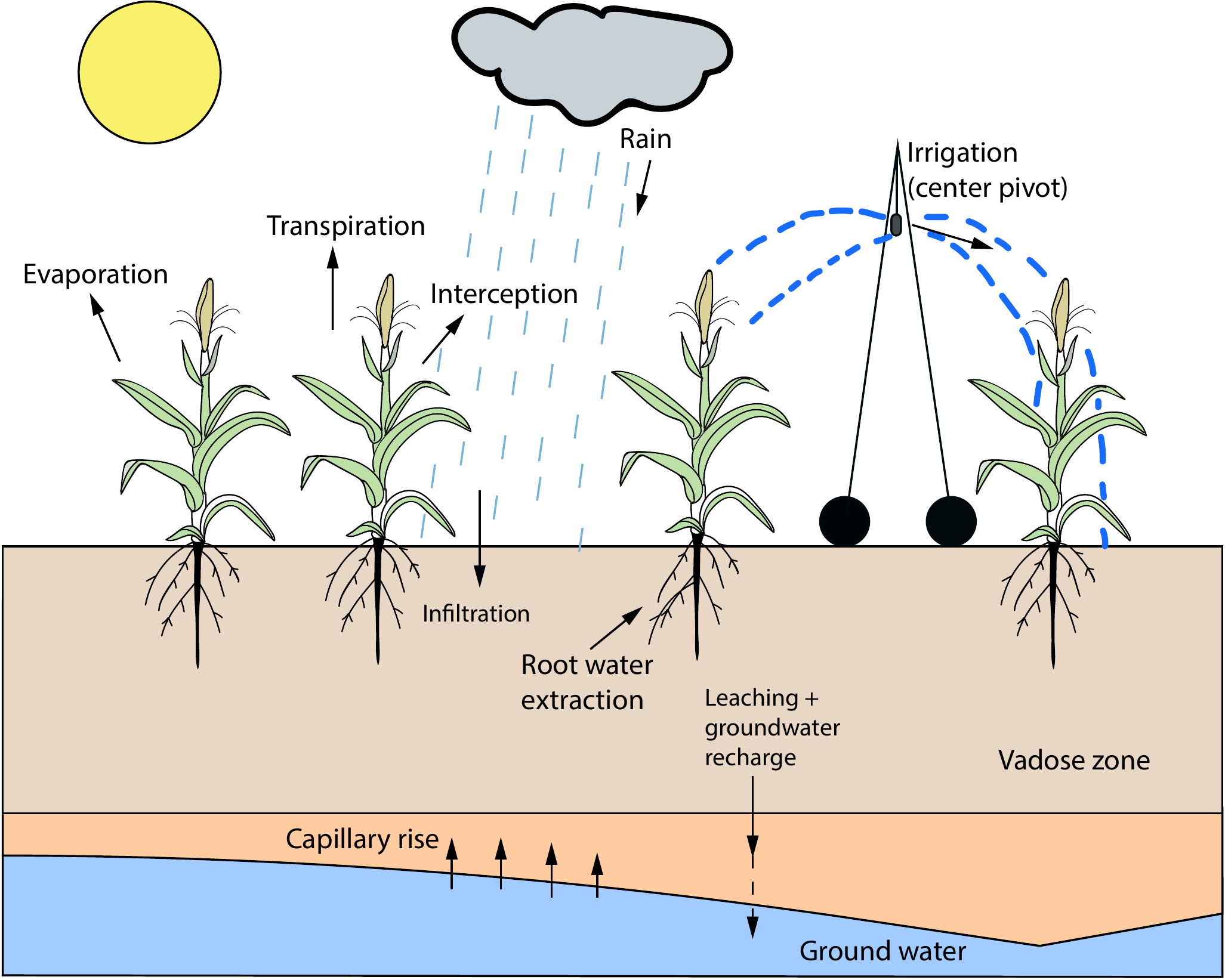}}
	\caption{An agro-hydrological system.}
	\label{fig:Polar_Agrohydrological}
\end{figure}

An agro-hydrological system characterizes the movement of water between the soil, crops, and the atmosphere. Figure \ref{fig:Polar_Agrohydrological} provides a simple illustration of an agro-hydrological system. 
In this system, water transport takes place primarily through irrigation, precipitation, evaporation, transpiration, infiltration, root water extraction, surface run-off, and drainage. The transport  of water in soil under the action of capillary and gravitational forces can be modeled using the Richards equation. Models that describe the dynamics of soil moisture, such as the Richards equation, are the most used tools for irrigation scheduling since the water available for uptake by crops can be inferred from soil moisture~\cite{majone2013wireless}. The Richards equation can be expressed in capillary pressure head form as:
\begin{equation}\label{eq:Richardseqn_general}
c(\psi)\frac{\partial \psi}{\partial t}=\nabla.~(K(\psi)\nabla~(\psi+z)) -S(\psi,z)
\end{equation}
In Equation (\ref{eq:Richardseqn_general}), $\psi[\text{L}]$ is the capillary pressure head, which describes the status of water in soil, $t[\text{T}]$ represents time, $z[\text{L}]$ is the spatial coordinate, $K(\psi)[\text{L}~\text{T}^{-1}]$ is the unsaturated hydraulic water conductivity, $ c(\psi)~[\text{L}^{-1}]$ is the capillary capacity. $K(\psi)$ and $ c(\psi)$ are parameterized by models of Maulem~\cite{mualem1976new} and van Genuchten~\cite{van1980closed}. $S\left(\psi,z\right)[\text{L}^3\text{L}^{-3}\text{T}^{-1}]$ denotes the sink term and it is expressed as:
\begin{equation}
S(\psi,z) = \alpha(\psi)\mathcal{R}\left(K^c, ET^0, z_r\right)
\end{equation}
$\alpha(\psi)[-]$ is a dimensionless stress water factor, $\mathcal{R}(\cdot)$ is the root water uptake model which is a function of the crop coefficient $K^c[-]$, the reference evapotranspiration $ET^0[\text{LT}^{-1}]$, and the rooting depth $z_r[\text{L}]$.

\subsection{Long short-term memory network}
\begin{figure}[H]
	\centering
	\centerline{\includegraphics[width=0.4\textwidth]{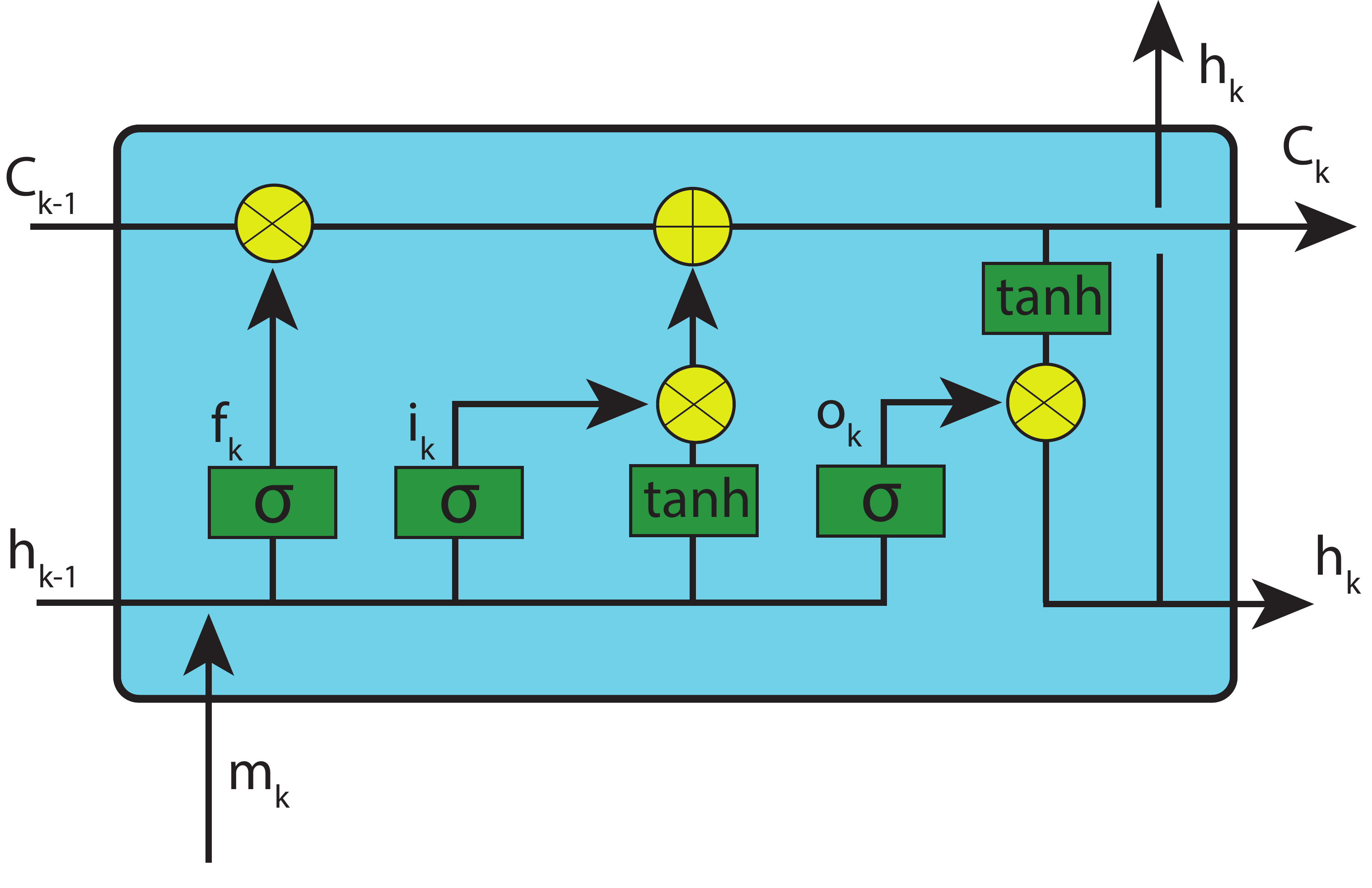}}
	\caption{The LSTM network memory block.}
	\label{fig:lstm_unit}
\end{figure}

LSTMs are a special kind of RNN that are capable of learning long-term dependencies. In addition to the hidden state ($h$) which is common to all RNNs, an LSTM memory block (Figure \ref{fig:lstm_unit}) is composed of a cell state ($C$), an input gate ($i$), an output gate ($o$), and a forget gate ($f$). The cell state is responsible for remembering information over arbitrary time intervals. The gates are made up of a sigmoid activation function ($\sigma$) and  a pointwise multiplication operation ($\otimes$). Intuitively, each of the three gates can be thought of as regulators of the information that is contained in the cell state.

Given an input sequence $m_k,~k=1,...,T$, where $T$ is the length of the input sequence, the LSTM evaluates the mapping of the inputs to the predicted output sequence $\hat{x}_k$ by looping through the following equations:
\begin{gather}
 i_k=\sigma(w_im_k + U_ih_{k-1} + b_i)\\
f_k=\sigma(w_fm_k + U_fh_{k-1} + b_f)\\
o_k=\sigma(w_om_k + U_oh_{k-1} + b_o)\\
\tilde{C}_k = \text{tanh}(w_cm_k + U_ch_{k-1} + b_c)\\ 
C_k = f_k\otimes C_{k-1} + i_k\otimes \tilde{C}_k\\
h_k = o_k\otimes \text{tanh}(C_k)\\
\hat{x}_k = w_yh_k+ b_y\label{eq:final}
\end{gather}
where $w_i~, w_f, w_o$ are the weights for the input, forget, and output gates to the input, respectively. $U_i,~U_f,~U_o$ are the matrices of the weights from the input, forget, and output gates to the hidden states, respectively. $b_i,~b_f,~b_o$ are the bias vectors associated with the input, forget, and output gates. The predicted state from the LSTM is calculated using Equation (\ref{eq:final}) where $w_y$ and $b_y$ denote the weight matrix and bias vector for the output, respectively.

\section{Methodology}
\subsection{Development of the LSTM model}\label{lstm_model}
In this section, we discuss how to develop an LSTM model for an agro-hydrological system using training data obtained from the Richards equation. Particularly, the development of the LSTM model includes the generation of the dataset, a specification of the model's framework, and the training process.

\subsubsection{Data generation}
In this work, we focus on infiltration processes in agro-hydrological systems. Infiltration is often assumed to be a one-dimensional (1D) process in the vertical direction~\cite{farthing2017numerical}; thus, the 1D Richards equation in the $z$-direction is used in this work. The 1D Richards equation is expressed as:
\begin{equation}\label{eq:RE_1D}
 c(\psi)\frac{\partial \psi}{\partial t} = \frac{\partial}{\partial z}\bigg[K(\psi)\bigg(\frac{\partial \psi  }{\partial z}+ 1\bigg)\bigg]-\alpha(\psi)\mathcal{R}\left(K^c, ET^0, z_r\right)
\end{equation}

\begin{remark}
	The 1D Richards equation that is adopted in this work is unable to account for horizontal variability in soils. Thus, in instances where there exists significant geospatial variation in the field under examination, higher dimensional versions of the Richards equation will be more suitable.  In heterogeneous soils, the 1D Richards equation can be invoked at multiple locations of the field. It is worth noting that this work adopts this approach when it comes to adapting the proposed scheduler to spatially variable fields. 
\end{remark}
 Equation (\ref{eq:RE_1D}) is solved numerically using the method of lines approach. The central difference scheme is used to approximate the spatial derivative. Implicit schemes, specifically the Backward Differentiation Formulas (BDFs), are used to approximate the time derivative. In order to solve for both saturated and unsaturated conditions, the numerical scheme is adapted such that only multiplication with $c(\psi)$ occurs~\cite{van2000numerical}. The final representation of the 1D Richards equation, after carrying out the temporal and spatial discretizations is expressed as:
\begin{multline}\label{eq:richards_eqn_discr}
c^{p+1}_{k}(\psi)\left(\frac{\psi^{p+1}_{k} - \psi^{p}_{k} }{\Delta t}\right) = \Bigg(\frac{1}{\Delta z_k}\bigg[K^p_{k+\frac{1}{2}}(\psi)\bigg(\frac{\psi^{p+1}_{k+1}-\psi^{p+1}_{k}}{\Delta z_N} +1\bigg)~~~~\\
-K^p_{k-\frac{1}{2}}(\psi)\bigg(\frac{\psi^{p+1}_{k}-\psi^{p+1}_{k-1}}{\Delta z_S}+1\bigg)\bigg]\Bigg)- \alpha(\psi)\mathcal{R}\left(K^c, ET^0, z_r\right)\Bigg]
\end{multline}
$k \in [1, N_z] $, $N_z$ is the number of nodes in the $z$-direction. $\Delta z_N= z_{k+1}-z_{k}$,  $\Delta z_S= z_{k}-z_{k-1}$, $\Delta z_k = \frac{1}{2}(z_{k+1}-z_{k-1})$, and  $K_{k\pm\frac{1}{2}}\left(\psi\right)\approx\frac{1}{2}(K\left(\psi_{k}\right)+K\left(\psi_{k\pm 1}\right))$. $p$ and $\Delta t$ represent the time level and time step, respectively.
Equation (\ref{eq:richards_eqn_discr}) is solved numerically for the following initial and boundary conditions:
\begin{align}
\psi_k(t=0)&= \psi^{\text{init}}\\
\frac{\partial (\psi+z)}{\partial z}\bigg|_{z=H_z}&=1\label{eq:botBC} \\
\frac{\partial \psi}{\partial z}\bigg|_{z=0}&=-1-\frac{u^{\text{irrig}}}{K(\psi)}\label{eq:topBC}
\end{align}
$H_z$ and $u^{\text{irrig}}$ [$\text{L}\text{T}^{-1}$] in Equations (\ref{eq:botBC}) and (\ref{eq:topBC})  represent the depth of the soil column and the irrigation rate, respectively. The depth dependent root water uptake model proposed by~\cite{feddes1982simulation} is used as the sink term in this work. Equation (\ref{eq:richards_eqn_discr}) together with the initial and boundary conditions is expressed in state space form as:
\begin{eqnarray}\label{eq:statespace}
x_{k+1}=\mathcal{F}(x_k,u_k) + \omega_k
\end{eqnarray}
where $x_k\in \mathbb{R}^{N_x}$ represents the state vector containing $N_x=N_z$ capillary pressure head values for the corresponding spatial nodes. $u_k$ represents the input vector containing the irrigation amount, precipitation, daily reference evapotranspiration, and the crop coefficient. $\omega_k$ is the model disturbance. 

Extensive open-loop simulations are conducted to generate a dataset that captures the soil water content dynamics for the state $x$ and the inputs $u$. Using randomly generated initial states $x_0$, Equation (\ref{eq:statespace}) is solved for randomly generated inputs in order to obtain a large number of state trajectories. In order to ensure a small temporal truncation error, the open-loop simulations are performed with a  small time step size. It should also be noted that, model uncertainty (noise) is included in the open-loop simulations. This leads to a noisy dataset which can improve the generalization ability and the robustness of the LSTM. Finally, the time-series data obtained from the open-loop simulations are partitioned into training, validation, and test datasets.

\begin{remark}
	The numerical solution of the Richards equation under extremely dry and wet conditions is unreliable\cite{farthing2017numerical}. Thus, during the open-loop simulation stage, it is more suitable to consider initial conditions, inputs (irrigation amount, reference evapotranspiration), and boundary conditions that guarantee a successful and reliable numerical solution of the Richards equation. In this paper, for example, very high reference evapotranspiration values which often result in extremely dry conditions are avoided. 
\end{remark}

\subsubsection{Proposed LSTM model and model training}\label{framework}
For irrigation scheduling purposes, it suffices to focus on the soil moisture dynamics in the root zone of the investigated soil column. Thus, we propose an LSTM model that predicts the root zone capillary pressure head in an agro-hydrological system. Specifically, the LSTM is trained to predict the the one-day-ahead root zone capillary pressure head $x_{t+1}$ using the present and the past root zone capillary pressure head $x(t=0,...,l)$, the irrigation amount inputs $u^{\text{irrig}} (t=0,...,l)$, the rain inputs $r (t=0,...,l)$, the crop coefficient inputs $K^c (t=0,...,l)$, and the reference evapotranspiration inputs $ET^0 (t=0,...,l)$. $l$ is the time lag used for the model development, whose value is determined through experimentation. The proposed LSTM model can thus be described as a multiple input and single output system. In order to realize the LSTM model, the states outside the root zone are discarded from the datasets and the resulting datasets are resampled to a time frame of 1 day. In addition to approximating the complex 1D Richards equation, the proposed LSTM model can also be thought of as a reduced model since it has fewer states (lower order) compared to Equation (\ref{eq:statespace}), which is known to have a higher order.

Prior to training the LSTM model, the datasets are normalized to rescale the input and output variables. The LSTM model is trained with the Keras Deep Learning Library in Python. The optimal number of layers and LSTM units are determined through experimentation. During the training process, an optimization problem which minimizes the modeling error is solved using an adaptive moment estimation algorithm (i.e. Adam in Keras). The mean squared error between the predicted states and the actual states in the training dataset is chosen as the loss function for the optimization problem. To prevent over-fitting of the LSTM model, the training processes is terminated when the error in the validation stops decreasing.

\begin{remark}
	Though the proposed LSTM model is identified with simulated data from the Richards equation, it can also be developed with field data obtained from soil moisture and climate monitoring networks. After applying some data cleaning and pre-processing steps to the field data, the steps outlined in section \ref{framework} can be applied to the resulting dataset in a straightforward manner. 
\end{remark}

\subsection{Scheduler design - mixed-integer MPC}\label{original_formulation}
\begin{figure}[t]
\centering
\includegraphics[width=0.9\textwidth]{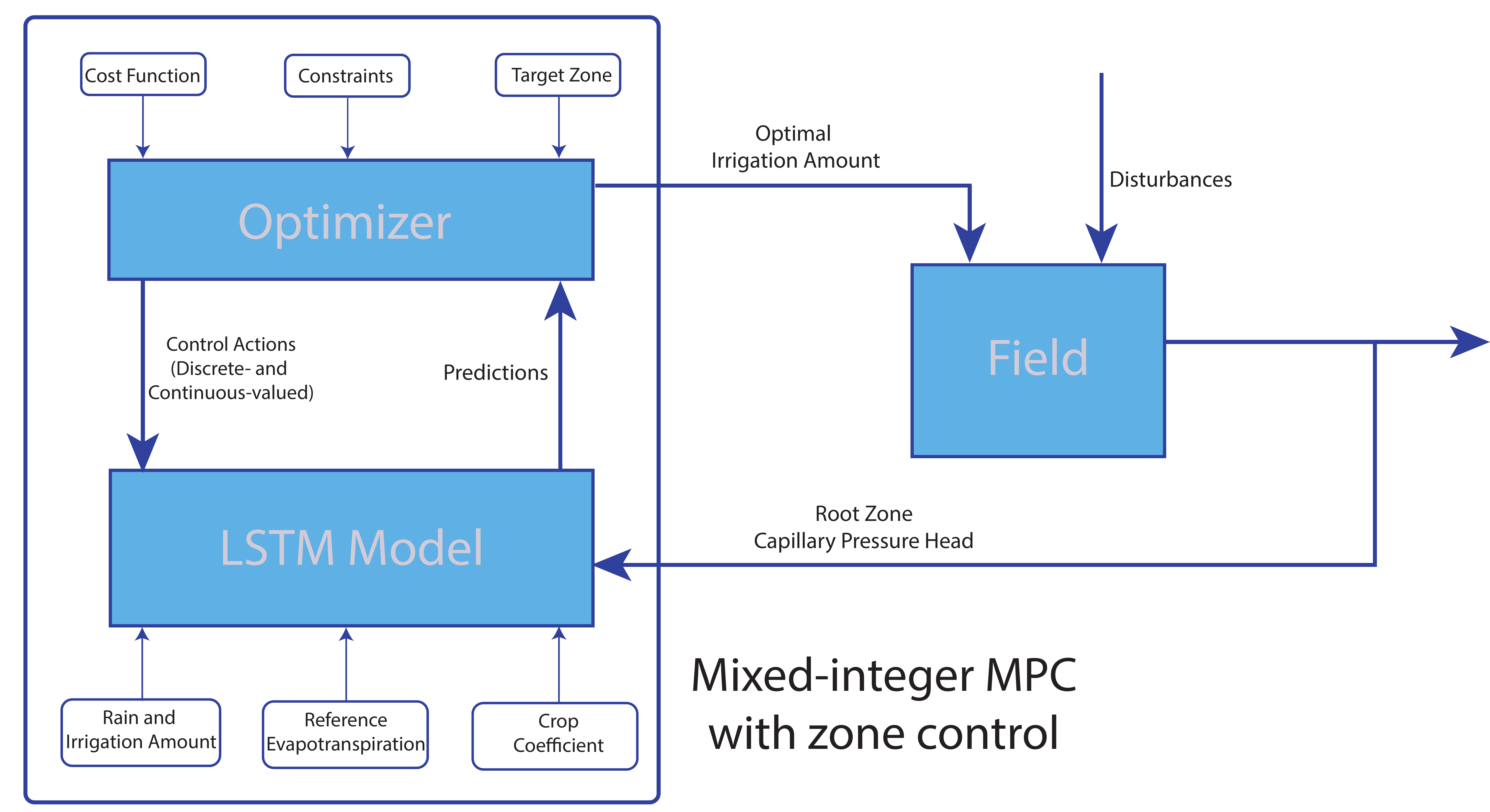}
\caption{A block diagram of the proposed irrigation scheduler.}
\label{fig:scheduler} 
\end{figure}

 The proposed scheduler (Figure \ref{fig:scheduler}) is designed in the mixed-integer MPC with zone control framework. The  scheduler considers a prediction horizon of up to a few weeks and its primary objective is to ensure optimal water uptake in crops while minimizing the total water consumption and the irrigation cost. In this design, the scheduler ensures optimal water uptake in crops by maintaining the root zone capillary pressure head within a target zone. The integer (binary) variable embedded in this design encodes the daily discrete (yes/no) irrigation decision. Using past weather data, daily weather forecast, the  root zone capillary  pressure head measurement, and the identified LSTM model, the scheduler  prescribes the daily discrete irrigation decision and the daily irrigation amount that achieve its primary objective. Additionally, the soft constraint approach is used to realize the zone control in this design. In this approach, slack variables are introduced in the formulation to relax the limits (bounds) of the target zone. At the same time, the slack variables are included in the objective function that is to be minimized.  For day $d$ and a fixed prediction horizon of $N$, the scheduler is formulated as follows:
 \begin{subequations}
\begin{align}
\min_{\bm{x},~\bm{\bar{\epsilon}},~\bm{\underline{\epsilon}},~\bm{u}^{\text{irrig}},~\bm{c}} ~~~~~&\sum_{k=d+1}^{d+N}\left[\bar{Q}\bar{\epsilon}^2_k + \underline{Q}\underline{\epsilon}^2_k \right] + \sum_{k=d}^{d+N-1}R_cc_k + \sum_{k=d}^{d+N-1}R_uu_k^{\text{irrig}} \label{eq:obj} \\
\notag
  \qquad \textrm{subject to }\\ \notag
   &x_{k+1} = \mathcal{F}_{\text{LSTM}}(\{x\}_{k-l}^{k}, \{K^c\}_{k-l}^{k}, \{ET^0\}_{k-l}^{k}, \{u^{\text{irrig}}\}_{k-l}^{k},\theta), \\ 
   &\hspace{5cm} k\in [d,d+N-1] \label{eq:cons1}\\
   &x_d  = x(d) \label{eq:cons2} \\
   &\underline{\nu} - \underline{\epsilon}_k \leq x_k \leq \bar{\nu} + \bar{\epsilon}_k, \hspace{1.4cm} k\in [d+1,d+N] \label{eq:cons3} \\
   &c_k\underline{u}^{\text{irrig}} \leq u_k^{\text{irrig}} \leq c_k\bar{u}^{\text{irrig}}, \hspace{0.8cm} k\in [d,d+N-1] \label{eq:cons4} \\
  &c_k= \{0, 1\}, \hspace{3.1cm} k\in [d,d+N-1] \label{eq:cons5}\\ 
&\underline{\epsilon}_k \geq 0, \quad \bar{\epsilon}_k\geq 0, \hspace{2.3cm} k\in [d+1,d+N] \label{eq:cons6} 
\end{align}
\end{subequations}
where $k \in \mathbb{Z}^+$, $\bm{x}\coloneqq [ x_d, x_{d+1},...,x_{d+N}]$, $\bm{\bar{\epsilon}}\coloneqq [ \bar{\epsilon}_{d+1}, \bar{\epsilon}_{d+2},...,\bar{\epsilon}_{d+N}]$, $\bm{\underline{\epsilon}}\coloneqq [ \underline{\epsilon}_{d+1}, \underline{\epsilon}_{d+2},...,\underline{\epsilon}_{d+N}]$, $\bm{c}\coloneqq [ c_{d}, c_{d+1},...,c_{d+N-1}]$, $\bm{u}^{\text{irrig}}\coloneqq [ u_{d}^{\text{irrig}}, u_{d+1}^{\text{irrig}},...,u_{d+N-1}^{\text{irrig}}]$, and $\{\gamma\}_{k-l}^{k} \coloneqq [ \gamma_{k-l}, \gamma_{k-l-1},\gamma_{k-l-2},..,\gamma_{k}]$ where $\gamma\in[x,K^c, ET^0, u^{\text{irrig}}]$. 
$\underline{\epsilon}_k$ and $\bar{\epsilon}_k$~(\ref{eq:obj}, \ref{eq:cons6}) are nonnegative slack variables that are introduced to relax the target zone $(\underline{\nu}_k,~\bar{\nu}_k)$ in~(\ref{eq:cons3}). $\underline{Q}$ and $\bar{Q}$ are the per-unit costs associated with the violation of the lower and upper bounds of the target zone, respectively. $R_c$ is the fixed cost associated with the operation of the irrigation implementing system, and $R_u$ is the per-unit cost of the irrigation amount $u^{\text{irrig}}$. The binary variable ($c$) encodes the daily discrete irrigation decision. The cost function~(\ref{eq:obj}) incorporates the objectives of maintaining the root zone capillary pressure head in a target zone in order to ensure optimal water uptake in crops by minimizing the violation of the target zone $\sum_{k=d+1}^{d+N}\left[\bar{Q}\bar{\epsilon}^2_k + \underline{Q}\underline{\epsilon}^2_k \right]$, minimizing the irrigation cost $\sum_{k=d}^{d+N-1}R_cc_k$, and minimizing the irrigation amount $\sum_{k=d}^{d+N-1}R_uu_k^{\text{irrig}}$. Constraint~(\ref{eq:cons1}) corresponds to the LSTM model of the root zone capillary pressure head. The initial state is assumed to be measured and it is represented with Constraint~(\ref{eq:cons2}). Constraint~(\ref{eq:cons4}) is the amount of water that can be supplied during the irrigation event on day $k$. When the irrigation decision on day $k$ is ``a no decision'' ($c_k=0$), Constraint~(\ref{eq:cons4}) specifies that the irrigation amount must necessarily be 0. On the other hand, when the irrigation decision on a particular day is ``a yes decision'' ($c_k=1$), this constraint states that the  prescribed irrigation amount must be at least equal to $\underline{u}^{\text{irrig}}$ and must be no larger than $\bar{u}^{\text{irrig}}$. The solution to the optimization problem (Equations (\ref{eq:obj}) - (\ref{eq:cons6})) is a sequence of predicted states $\bm{x}$, optimal slack variables ($\bm{\bar{\epsilon}}$, $\bm{\underline{\epsilon}}$), optimal irrigation decisions ($\bm{c}$), and optimal irrigation amounts ($\bm{u}$).

\begin{remark}
 Crop yield is related to the water stress factor $\mathcal{K}_s$, a dimensionless variable that characterizes the suppression of root water uptake due to drought and poor aeration. When the soil moisture lies between field capacity and the wilting point, $\mathcal{K}_s$ assumes a value of 1, which signifies the absence of water stress~\cite{feddes1982simulation}. Thus, the objective of maintaining the root zone capillary pressure head between field capacity and wilting point can be thought of as an indirect way of maximizing crop yield.
\end{remark}

\subsection{Scheduler design - sigmoid approximation} \label{sigmoid_formulation}
It is desirable to develop modifications to mixed-integer problems so as to ensure that they can be executed in real-time. This is necessary because mixed-integer programming belongs to the class of $\mathcal{NP-}$complete problems and thus, can require extensive computation time and resources for problems with many integer variables. To this end, a heuristic method that approximates the binary variable in the mixed-integer formulation, which was originally applied to transmission expansion problems~\cite{mazzini2018minimisation} and active power losses minimization problems~\cite{de2005transmission}, is employed. Specifically, the binary variable in Equations (\ref{eq:obj}) - (\ref{eq:cons6}) is replaced with a sigmoid function $\omega(r)$ which is defined as:
\begin{equation}\label{eq:sigmoid}
\omega(r) = \frac{1}{1+e^{-\beta r}}
\end{equation}
where $\beta$ is the slope of the sigmoid function and the argument $r$ is a real number. The inclusion of Equation (\ref{eq:sigmoid}) in Equations (\ref{eq:obj}) - (\ref{eq:cons6}) results in the modified problem:
\begin{subequations}
\begin{align}
\min_{\bm{x},~\bm{\bar{\epsilon}},~\bm{\underline{\epsilon}},~\bm{u}^{\text{irrig}},~\bm{r}} ~~~~~&\sum_{k=d+1}^{d+N}\left[\bar{Q}\bar{\epsilon}^2_k + \underline{Q}\underline{\epsilon}^2_k \right] + \sum_{k=d}^{d+N-1}R_c\omega(r_k) + \sum_{k=d}^{d+N-1}R_uu_k^{\text{irrig}} \label{eq:obj_sig} \\
\notag
  \qquad \textrm{subject to }\\ \notag
  & x_{k+1} = \mathcal{F}_{\text{LSTM}}(\{x\}_{k-l}^{k}, \{K^c\}_{k-l}^{k}, \{ET^0\}_{k-l}^{k}, \{u^{\text{irrig}}\}_{k-l}^{k},\theta),\\ 
  & \hspace{5.9cm} k\in [d,d+N-1] \label{eq:cons1_sig}\\
  & x_d = x(d) \label{eq:cons2_sig} \\
 & \underline{\nu} - \underline{\epsilon}_k \leq x_k \leq \bar{\nu} + \bar{\epsilon}_k, \hspace{2.25cm} k\in [d+1,d+N] \label{eq:cons3_sig} \\
 &\omega(r_k)\underline{u}^{\text{irrig}} \leq u_k^{\text{irrig}} \leq \omega(r_k)\bar{u}^{\text{irrig}},\hspace{0.53cm} k\in [d,d+N-1] \label{eq:cons4_sig} \\
 &r_{\text{min}}\leq r_k\leq r_{\text{max}}, \hspace{2.96cm} k\in [d,d+N-1] \label{eq:cons5_sig}\\ 
& \underline{\epsilon}_k \geq 0,  \quad \bar{\epsilon}_k\geq 0, \hspace{3.15cm} k\in [d+1,d+N] \label{eq:cons6_sig} 
\end{align}
\end{subequations}
where $\bm{r}\coloneqq [ r_d, r_{d+1},...,r_{d+N-1}]$. The modified problem is an nonlinear program (NLP) and can thus be solved with a suitable NLP algorithm.

\subsubsection{Selection of $\beta$}
The sigmoid function converges to binary elements (0 or 1) for higher values of its slope $\beta$. However, the use of very large $\beta$ values often results in an ill-conditioned optimization. To handle this issue, an algorithm is proposed to improve the convergence of the sigmoid function to binary elements while reducing ill-conditioning issues. This process involves successively solving Equations (\ref{eq:obj_sig}) - (\ref{eq:cons6_sig}) for increasing values of $\beta$ (by a factor of $\tau$ in Algorithm \ref{alg:cap}) until a predetermined convergence criterion is met. This predefined criterion, for the $i^{\text{th}}$ evaluation of Equations (\ref{eq:obj_sig}) - (\ref{eq:cons6_sig}),  can be mathematically expressed as:
\begin{equation}\label{eq:criterion}
\Vert \omega(\bm{r}^i) - \bm{c}\Vert _2 \leq \zeta
\end{equation} 
where $\bm{c}$ is a vector of binary elements. Each element of $\bm{c}$ corresponds to the nearest binary value of each element of $\omega(\bm{r}^i)$. $\zeta$ represents the convergence tolerance. The detailed steps are described in Algorithm \ref{alg:cap}.
\begin{algorithm}
\caption{Algorithm for approximating $\bm{c}$ with $\omega(\bm{r})$}\label{alg:cap}
\begin{algorithmic}
\Require $x(0)$, $\bm{x}^0$, $\bm{r}^0$, ($\bm{u}^{\text{irrig}})^0$, $\zeta$, $\beta^0$, $\tau$ 
\State $x_{1} \gets x(0)$
\State $\bm{x}_{\text{guess}} \gets \bm{x}^0$
\State $\bm{r}_{\text{guess}} \gets \bm{r}^0$
\State ($\bm{u}^{\text{irrig}})_{\text{guess}} \gets (\bm{u}^{\text{irrig}})^0$
 \State $\beta \gets \beta^0$
\State $i \gets 0$
\While{$ \Vert \omega(\bm{r}^i) - \bm{c} \Vert _2 > \zeta$}\\
Solve Equations (\ref{eq:obj_sig})-(\ref{eq:cons6_sig}) for $\beta ^i$
\State $\bm{x}_{\text{guess}} \gets \bm{x}^i$
\State $\bm{r}_{\text{guess}} \gets \bm{r}^i$
\State ($\bm{u}^{\text{irrig}})_{\text{guess}} \gets (\bm{u}^{\text{irrig}})^i$
\State $\beta^{i+1} \gets \tau \beta^i$
\State $i \gets i+1$
\EndWhile
\end{algorithmic}
\end{algorithm}

\subsection{Scheduler design for spatially variable fields} \label{spatially_variable_formulation}
Due to biological, chemical, and physical processes that occur in agro-hydrological systems, high spatial variability is commonly found in soils. Consequently, irrigation scheduling that assumes a uniform field (uniform water application rate) is not the most effective approach and thus, the scheduler formulation described in section \ref{original_formulation} is not suitable for irrigation scheduling in spatially variable fields. It is necessary to characterize the within-field spatial variability during the determination of irrigation schedules in order to ensure improved water use efficiency and increased crop productivity. The basic approach to spatially variable irrigation scheduling is to delineate a spatially variable field into distinct management zones~\cite{moral2010delineationmoral2010delineation,king2005distributed}. Management zones are sub-field areas with homogeneous properties that are known to affect the yield of crops. A field can be delineated into management zones based on soil texture, elevation, topography, experience, drainage, among others. The irrigation timing and the water application depth can then be optimized for each distinct management zone. 
Motivated by the above, it is important to adapt the proposed scheduler to irrigation scheduling in spatially variable fields. To this end, we employ the following steps:
\begin{enumerate}
\item The spatially variable field under consideration is divided into definite management zones. The delineation of the field into management zones may be done based on soil texture, elevation, drainage, among others.
\item An LSTM model is then trained for each management zone using simulated data from the 1D Richards equation, according to the steps outlined in section~\ref{lstm_model}.
\item Irrigation can then be scheduled using the root zone capillary pressure head measurement for each management zone. It is worth noting that all areas within a given management zone receive the same irrigation amount. However, the irrigation amount may vary form one management zone to another. 
\end{enumerate}
In the context of spatially variable irrigation scheduling, we propose separate formulations that are applicable to small-scale and large-scale fields. The detailed formulations  are described in the sequel.

\subsubsection{Spatially variable irrigation scheduling for small-scale fields}
In this paper, a spatially variable field with distinct management zones is considered as a small-scale field if the irrigation implementing equipment is able to irrigate all the management zones in one day. For such fields, in principle, the irrigation implementing equipment can irrigate all the management zones on each day within the prediction horizon. Irrigation scheduling in small-scale spatially variable fields should thus reflect this observation in order to avoid over and under irrigation. For day $d$, $M$ management zones, and a prediction horizon $N$, the scheduler for a small-scale spatially variable field can be formulated as follows:
\begin{subequations}
\begin{align}\label{eq:obj_sv}
\min_{\bm{X, ~\bar{E}, ~\underline{E},~ U^{\text{irrig}},~ c}} ~~~~~& \sum_{j=1}^M \sum_{k=d+1}^{d+N}\left[\bar{Q}(j)\bar{\epsilon}^2_{jk} + \underline{Q}(j)\underline{\epsilon}^2_{jk} \right] + R_c\sum_{k=d}^{d+N-1}c_{k} +  R_u\sum_{j=1}^M\sum_{k=d}^{d+N-1}u_{jk}^{\text{irrig}}\\ 
\notag
  \qquad \textrm{subject to }\\ \notag
& x_{j(k+1)} = \mathcal{F}_{\text{LSTM}}(\{x_j\}_{k-l}^{k}, \{K^c_j\}_{k-l}^{k}, \{ET^0\}_{k-l}^{k}, \{u_j^{\text{irrig}}\}_{k-l}^{k},\theta),  \notag \\
& \hspace{6.15cm} j\in \tilde{M}, ~k\in [d,d+N-1]\\
&x_{jd} = x_j(d), \hspace{4.1cm} j\in \tilde{M}~\\
&\underline{\nu}_j - \underline{\epsilon}_{jk} \leq x_{jk}\leq \bar{\nu}_j + \bar{\epsilon}_{jk}, \hspace{1.9cm}j\in \tilde{M} ,~k \in [d+1,d+N]\\
&c_{k} \underline{u}_{j}^{\text{irrig}}\leq u_{jk}\leq c_{k} \bar{u}_{j}^{\text{irrig}}, \hspace{2.3cm} j\in \tilde{M} ,~k \in [d,d+N-1]\\
&c_{k}\in \{0, 1\}, \hspace{5.75cm} k \in [d,d+N-1]\\
&\underline{\epsilon}_{jk} \geq 0, \quad \bar{\epsilon}_{jk}\geq 0, \hspace{3.25cm}j\in \tilde{M} ,~k \in [d+1,d+N] \label{eq:lastconst_sv}
\end{align}
\end{subequations}
where $j \in \mathbb{Z}^+$, $k \in \mathbb{Z}^+$, $\bm{X} \coloneqq [\bm{x}_1,\bm{x}_2,....,\bm{x}_M ]$, $\bm{\bar{E}} \coloneqq [\bm{\bar{\epsilon}}_1,\bm{\bar{\epsilon}}_2,....,\bm{\bar{\epsilon}}_M ]$, $\bm{\underline{E}} \coloneqq [\bm{\underline{\epsilon}}_1,\bm{\underline{\epsilon}}_2,....,\bm{\underline{\epsilon}}_M ]$, and $\bm{U}^{\text{irrig}} \coloneqq [\bm{u}_1^{\text{irrig}},\bm{u}_2^{\text{irrig}},....,\bm{u}_M^{\text{irrig}} ]$. $\tilde{M}$ is a closed set which contains the $[j_1, j_2, ..., j_M]$ positional indices of the $M$ management zones.

\subsubsection{Spatially variable irrigation scheduling for large-scale fields}
In this paper, a spatially variable field with $M$ distinct management zones is considered as a large-scale field if the irrigation implementing equipment takes more than one day to irrigate the entire field. Specifically, we consider a scenario where the irrigation implementing equipment spends one day irrigating each management zone and thus spends $M$ days irrigating the entire field.
 For such fields, depending on the order in which the irrigation implementing equipment irrigates the management zones during a single irrigation cycle, each management zone can be irrigated on specific days within the prediction horizon.
  This scenario is best illustrated with a concrete example. Assuming that the spatially variable field under consideration consists of 2 management zones ($M_1, M_2$). Also, supposing that the irrigation implementing equipment starts each irrigation cycle from $M_1$. Then for a prediction horizon of 14 days, the irrigation implementing equipment completes its irrigation cycle in 2 days and can thus be operated 7 times within the prediction horizon. Additionally, $M_1$ can only be irrigated on days 1, 3, 5, 7,...., 13 while $M_2$ can only be irrigated on days 2, 4, 6,...., 14. The determination of irrigation schedules for such a field should be made to reflect this arrangement in order to ensure optimal uptake of water by crops.
Before introducing the formulation for an arbitrary number of management zones and an arbitrary prediction horizon, for the sake of simplicity and with no loss of generality, we assume that the order in which the irrigation implementing equipment irrigates the entire field coincides with the indices assigned to the $M$ management zones. For day $d$, management zones $M$, and a prediction horizon $N$, the scheduler for a large-scale spatially variable field can be formulated as follows:
\begin{subequations}
\begin{align}\label{eq:obj_sv_bg}
\min_{\bm{X, ~\bar{E}, ~\underline{E},~ U^{\text{irrig}},~ c}} ~~~~~& \sum_{j=1}^M \sum_{k=d+1}^{d+N}\left[\bar{Q}(j)\bar{\epsilon}^2_{jk} + \underline{Q}(j)\underline{\epsilon}^2_{jk} \right] + R_c\sum_{n=1}^{N_c}c_{n} +  R_u\sum_{j=1}^M\sum_{k=d}^{d+N-1}u_{jk}^{\text{irrig}}\\ 
\notag
  \qquad \textrm{subject to }\\ \notag
& x_{j(k+1)} = \mathcal{F}_{\text{LSTM}}(\{x_j\}_{k-l}^{k}, \{K^c_j\}_{k-l}^{k}, \{ET^0\}_{k-l}^{k}, \{u_j^{\text{irrig}}\}_{k-l}^{k},\theta),  \notag \\
& \hspace{6.05cm} j\in \tilde{M}, ~k\in [d,d+N-1]\\
&x_{jd} = x_j(d), \hspace{4cm} j\in \tilde{M}~\\
&\underline{\nu}_j - \underline{\epsilon}_{jk} \leq x_{jk}\leq \bar{\nu}_j + \bar{\epsilon}_{jk}, \hspace{1.75cm}j\in \tilde{M} ,~k \in [d+1,d+N]\\
\notag \\ 
&\begin{cases}
c_{n} \underline{u}_{j_1}^{\text{irrig}}\leq u_{j_1k}\leq c_{n} \bar{u}_{j_1}^{\text{irrig}}, &k = \left\{ Mi + j_1: i \in [0,N_c)\right\} , n \in[1, N_c]\\
u_{jk} = 0 & j\in \tilde{M}\setminus j_1,~k = \left\{ Mi + j_1: i \in [0,N_c)\right\}\label{eq:loc1}
\end{cases}\\
\notag \\
&\begin{cases}
c_{n} \underline{u}_{j_2}^{\text{irrig}}\leq u_{j_2k}\leq c_{n} \bar{u}_{j_2}^{\text{irrig}}, &k =\left\{ Mi + j_2: i \in [0,N_c)\right\} , n \in[1, N_c]\\
u_{jk} = 0 & j\in \tilde{M}\setminus j_2,~k = \left\{ Mi + j_2: i \in [0,N_c)\right\}
\end{cases}\\
\notag
&\hspace{3cm}\vdots\\
&\begin{cases}
c_{n} \underline{u}_{j_M}^{\text{irrig}}\leq u_{j_Mk}\leq c_{n} \bar{u}_{j_M}^{\text{irrig}}, &k =\left\{ Mi + j_M: i \in [0,N_c)\right\} , n \in[1, N_c]\\
u_{jk} = 0 & j\in \tilde{M}\setminus j_M,~k = \left\{ Mi + j_M: i \in [0,N_c)\right\}
\end{cases}\\\label{eq:loc2}
\notag \\ 
&c_{n}\in \{0, 1\}, \hspace{5.6cm} n \in [1, N_c]\\
&\underline{\epsilon}_{jk} \geq 0, \quad \bar{\epsilon}_{jk}\geq 0, \hspace{3cm}j\in \tilde{M} ,~k \in [d+1,d+N] \label{eq:lastconst_sv_bg}
\end{align}
\end{subequations}

where $i \in \mathbb{Z}^{0+}$, $j \in \mathbb{Z}^+$, $k \in \mathbb{Z}^+$, $N_c=\frac{N}{M}$,  $\tilde{M} = [j_1, j_2,...., j_M]$, and the notation $\tilde{M}\setminus j$ refers to the closed set $\tilde{M}$ excluding the element $j$. $N_c$ represents the number of irrigation cycles that can be performed within the prediction horizon under consideration. In this formulation, instead of defining a binary variable for each day within the prediction horizon, we define a binary variable for the $N_c$ cycles that can be performed within $N$. The closed set $ A = [M(n-1)+d, Mn+d-1]$ represents the days that make up the $n^{\text{th}}$ cycle. For the $j^{\text{th}}$ management zone, the specific days within $N$ on which the irrigation event can be performed is given by the sequence $ B =\left\{ Mi + j: i \in [0,N_c)\right\}$. The second part of Constraints (\ref{eq:loc1} - \ref{eq:loc2}) stipulates that  the irrigation amount must be 0 for the remaining $\tilde{M}\setminus j$  management zones on all the days in $B$. For the $n^{\text{th}}$ cycle, if $c_n = 1$, the first part of Constraints (\ref{eq:loc1} - \ref{eq:loc2}) specifies that
prescribed irrigation amount for the $j^{\text{th}}$ management zone must be at least equal to $\underline{u}_j^{\text{irrig}}$ and must be no larger than $\bar{u}_j^{\text{irrig}}$ on the day corresponding to $k=A\cap B$. On the other hand, if $c_n = 0$, this constraint specifies that the irrigation amount must necessarily be 0 for all the $M$ management zones on all the $k$ days in A. 

\section{Simulation case studies}
We design various simulation experiments to test the predictive performance of the proposed LSTM model framework, and the utility of the scheduler.

To evaluate the predictive capability of the proposed model framework, an LSTM model is first identified to predict the root zone capillary pressure head in a 0.6 m loamy-sand soil column. 
Prior to the data generation process, the soil column is spatially discretized into 30 equally spaced compartments and each compartment is initialized with a pressure head value of -180 mm. A sequence of open-loop simulations are performed for the resulting soil column with a full sweep through the irrigation amount, rain, reference evapotranspiration, and crop coefficient inputs. Specifically, in order to generate a dataset that adequately captures the pressure head dynamics in the investigated column, the irrigation amount, rain, reference evapotranspiration, and crop coefficient inputs are randomly chosen from the following respective ranges: [1.4 mm/day, 15.6 mm/day], [1.04 mm, 7.0 mm], [1.04 mm, 3.0 mm], and [0.50, 0.88]. Additionally, the soil column is irrigated in a periodic manner and the time interval between any two successive irrigation events is selected at random between 2 and 6 days. Process noise is considered in the open-loop simulations and it has a zero mean with a standard deviation of 5$\times$10$^{-3}$. Sampled data points representing the state trajectories are obtained from the open-loop simulations at a sampling time of 6 minutes and the simulations are performed for a period of 5,000 days. The pressure head value at  a depth of 0.5 m is chosen to characterize the pressure head in the root zone. Thus, the pressure head values at other spatial points are discarded and the resulting noisy time series data is resampled to a time frame of 1 day. An LSTM model is subsequently developed based on the resampled data set to predict the one-day-ahead root zone capillary pressure head using the Keras API. The LSTM model is designed to have two hidden layers. Each layer consists of 200 LSTM units and a sequence length of 5 days is used for the training. Consequently, the time lag $l$ associated with the inputs of the LSTM model is 4.

We employ three simulation case studies, namely Case Studies 1, 2, and 3 to demonstrate the utility of the scheduler designs outlined in sections \ref{original_formulation},  \ref{sigmoid_formulation}, and \ref{spatially_variable_formulation} of this paper, respectively. Case Study 1 is based on a uniform field composed of loamy soil. In this simulation experiment,  an LSTM model is first identified for a 0.6 m loamy soil column using the method outlined in section \ref{lstm_model}. The lower and upper bounds of the target zone are chosen as -820 mm and -690 mm, and the per unit costs associated with the violation of these zones are chosen as  $\bar{Q} =\underline{Q}$ = 9000. The fixed cost associated with the irrigation implementing equipment is selected as $R_c$ = 50 and the per unit cost of the irrigation amount is $R_u$ = 20. Two scenarios are investigated in Case Study 1: (i) Scenario 1: Absence of rain in the weather data; and (ii) Scenario 2: Presence of rain in the weather data. The mixed-integer optimization problem is solved using the BONMIN solver in CasADi (version 3.5.1), which is an open source library for numerical optimization.

Scenario 1 of Case Study 2 is employed to demonstrate the ability of the sigmoid approximation (section \ref{sigmoid_formulation}) to approximate the mixed-integer formulation of the scheduler while achieving remarkable computation speeds. The NLP resulting from this investigation is solved using the IPOPT solver in CasADi.

The proposed formulations for spatially variable irrigation scheduling are investigated in Case Studies 2 and 3. In Case Study 2, spatially variable irrigation scheduling is investigated for a small-scale field that is delineated into three management zones (MZs). The delineation of the field is done according to soil texture, which gives rise to a loam MZ, a loamy sand MZ, and a sand MZ. In this simulation, separate LSTM models are identified for each MZ using simulated data from the 1D Richards equation. The lower and upper bounds of the target pressure head range/zone are chosen as -850 mm and -690 mm for the loamy soil MZ, -340 mm and -180 mm for the loamy sandy soil MZ, and -290 mm and -160 mm for the sandy soil MZ. 

In Case study 3, spatially variable irrigation scheduling is investigated for a large-scale field that is delineated into two management zones (MZs). The delineation of the field in this simulation experiment is also done according to soil texture, which gives rise to a loam MZ and a loamy sand MZ. Separate LSTM models are identified for each MZ using simulated data from the 1D Richards equation and the lower and upper bounds of the target pressure head range/zone are  chosen as -850 mm and -690 mm for the loamy soil MZ, and -340 mm and -180 mm for the loamy sandy soil MZ. 

In Case Studies 2 and 3, the sigmoid function approximation versions of Equations (\ref{eq:obj_sv}) - (\ref{eq:lastconst_sv}) and Equations (\ref{eq:obj_sv_bg}) - (\ref{eq:lastconst_sv_bg}) are utilized, respectively. The resulting NLP is solved using the IPOPT solver in CasADi. The values of $R_c$, and $R_u$ that are used in Case Study 1 are adopted in these case studies. Similarly, the values of $\bar{Q}$ and $\underline{Q}$ that  are used in Case Study 1 are adopted for the loam MZ. For the loamy sand and sand MZs, $\bar{Q}$ and $\underline{Q}$ are chosen as 2000 and 9000, respectively.

\begin{figure}[H]
	\subfloat[Reference evapotranspiration.] {
		\begin{minipage}[c][0.75\width]{
				0.45\textwidth}\label{fig:et_information}
			\centering
			\includegraphics[width=1\textwidth]{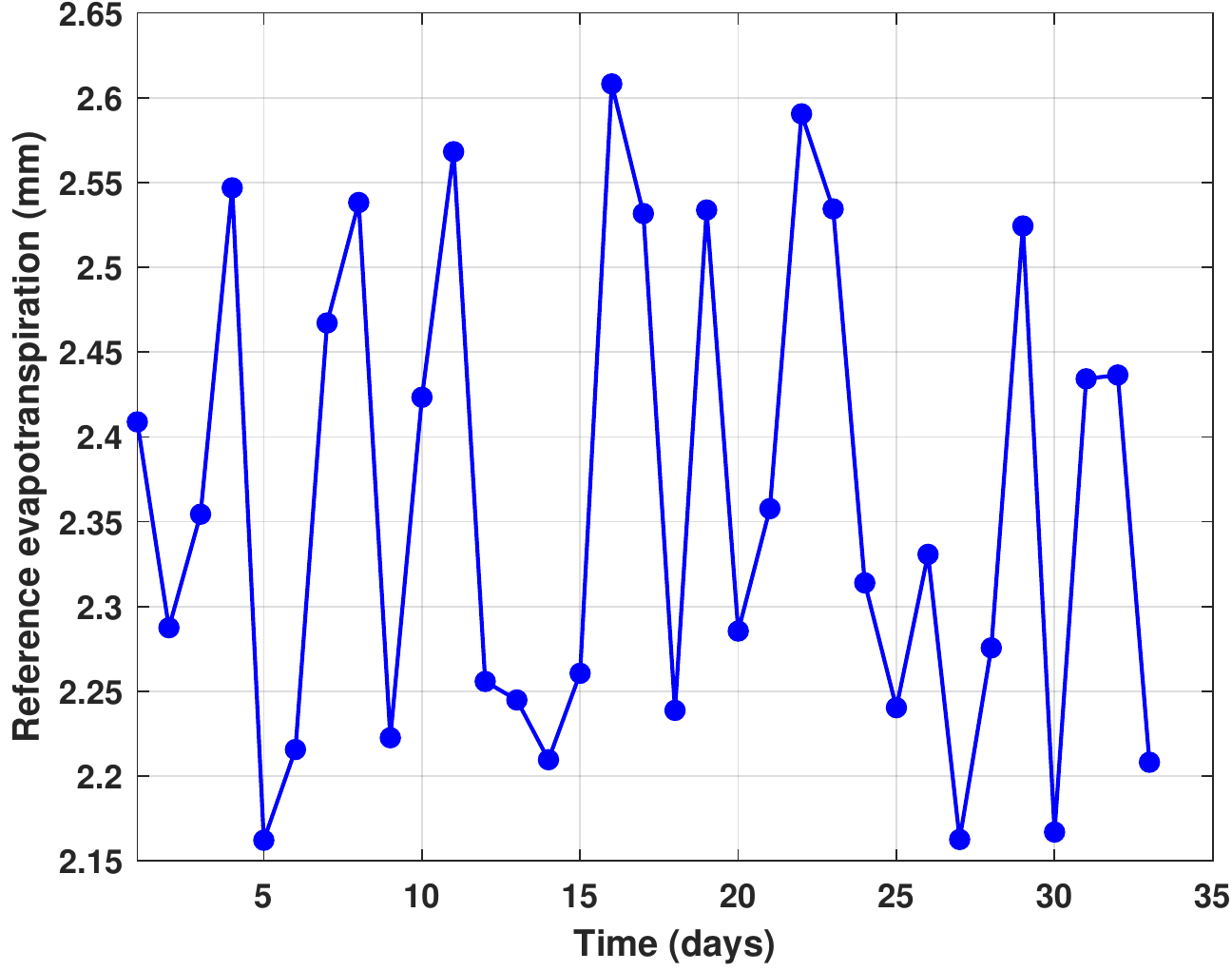}
	\end{minipage}}
	\hfill
	\subfloat[Rain information for Scenario 2 in Case Study 1.]{
		\begin{minipage}[c][0.75\width]{
				0.45\textwidth}\label{fig:rain_information}
			\centering
			\includegraphics[width=1\textwidth]{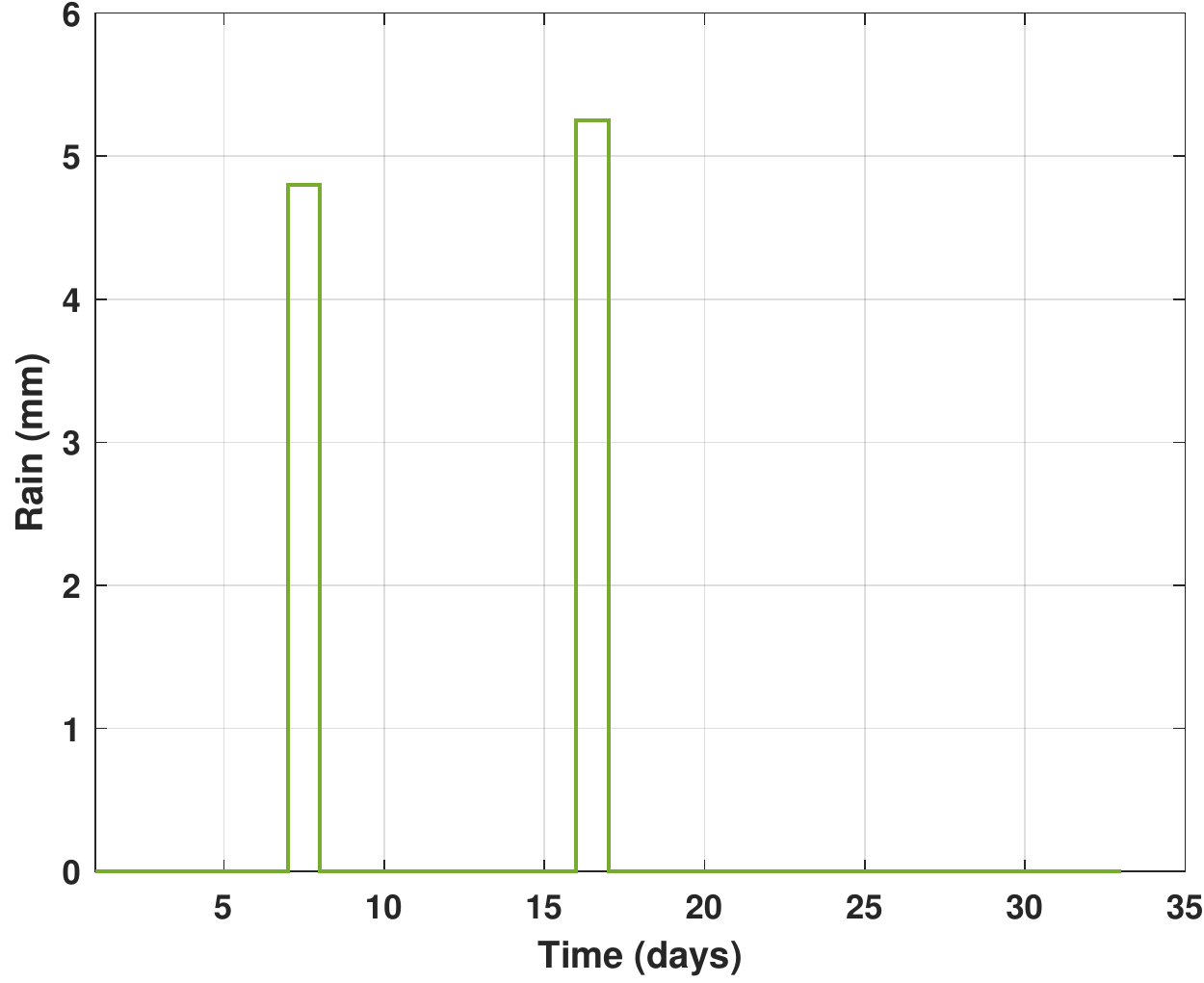}
	\end{minipage}}
	\caption{Weather data for the simulation case studies.}
\end{figure}

A prediction horizon of 14 days is used in all the simulations and the reference evapotranspiration data for the simulation period in all the three case studies is depicted in Figure \ref{fig:et_information}. The rain information that is used in Scenario 2 of Case Study 1 is depicted in Figure \ref{fig:rain_information}. A crop coefficient value of 0.5 is used in all the case studies. During the data generation step, for the various soil types considered in this work, the soil hydraulic parameters of the parametrized models of  $K(\psi)$ and $ c(\psi)$ are depicted in Table \ref{tbl:soil_pars}. In all the case studies, the scheduler is evaluated for  initial states of -754 mm in the loam MZ, -279 mm in the loamy sand MZ, and -262 mm in the sand MZ. Similarly, in all the case studies, past root zone capillary pressure head values of [-795 mm, -748 mm, -735 mm, -740 mm], [-173 mm, -212 mm, -239 mm, -260 mm], and [-225 mm, -239 mm, -251 mm, -258 mm] are used in the loam, loamy sand, and sand MZs, respectively. The lower and upper bounds of the target zone in all the three MZs lie within the field capacity and the permanent wilting points of loam, loamy sand, and sandy soils. 


\begin{table}[t]
	\caption{The hydraulic parameters of soil types considered in the simulation case studies.}
	\small 
	\centering
	\begin{tabular}{cccccc}
		\hline
		{ Soil Type }& {$K_{s}$ (m/s)} &{$\theta_{s}$ $(\text{m}^{3}/\text{m}^{3})$}&{$\theta_{r}$ $(\text{m}^{3}/\text{m}^{3})$}& {$\alpha$ (1/m)}& {$n$ (-)}\\
		\hline
		Loam & $2.889\times 10^{-6}$  & 0.430 & 0.078&3.600&1.56\\
		Loamy Sand & $4.053\times 10^{-5}$  & 0.410 & 0.057&12.40&2.28\\
		Sand & $8.250\times 10^{-5}$  & 0.430 & 0.045&14.50&2.68\\
		\hline
	\end{tabular} \label{tbl:soil_pars}
\end{table}

\section{Results and discussion}

\subsection{Predictive capability of the LSTM model}
\begin{figure}[H]
	\centering
	\includegraphics[width=0.7\textwidth]{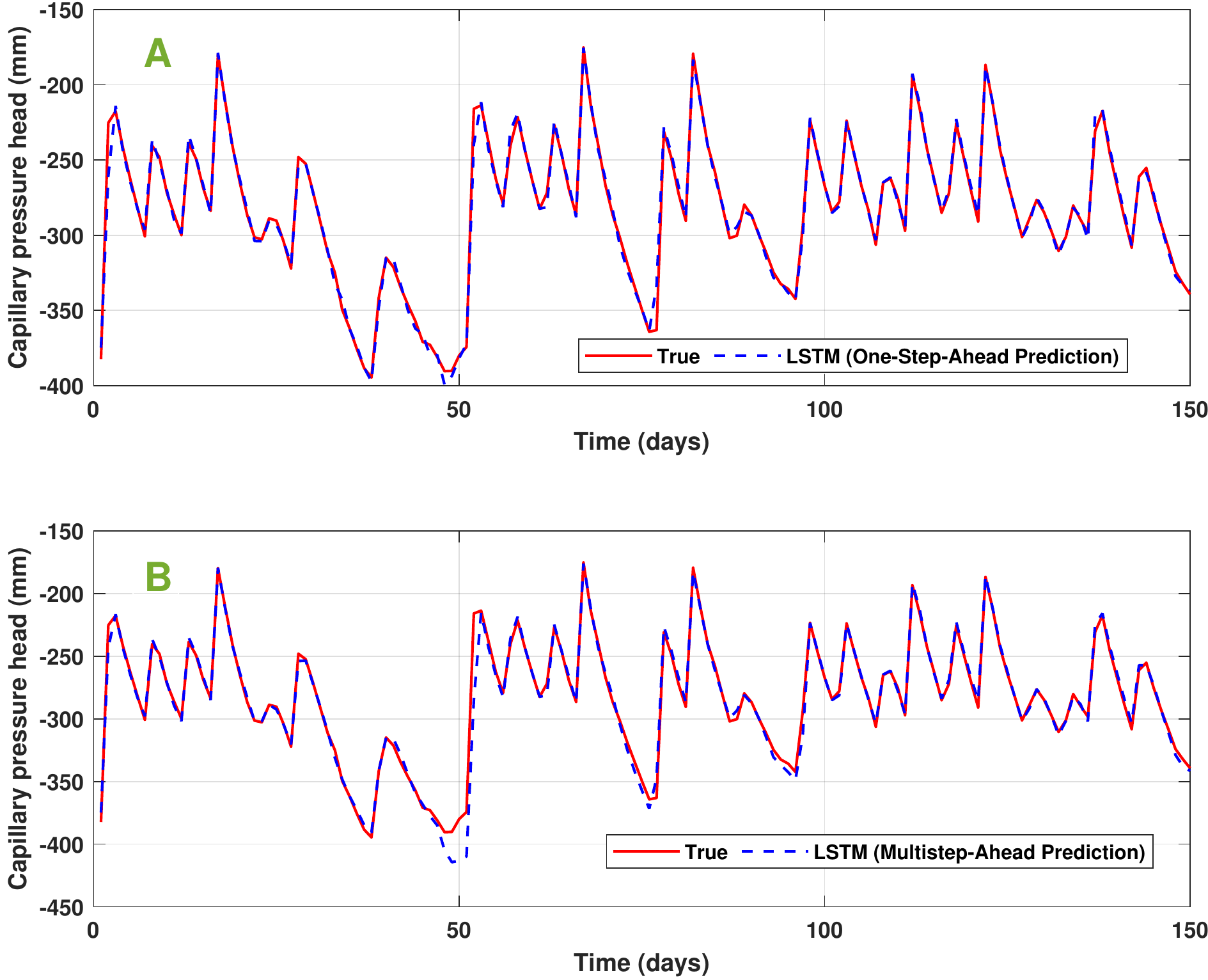}
	\caption{Actual root zone pressure head (red solid line) and the predicted root zone pressure head (blue dash-dot) using the test dataset, (\textbf{A}) One-day-ahead prediction, and (\textbf{B}) Multistep-ahead prediction.}
	\label{fig:prediction_performance}
\end{figure}

In Figure \ref{fig:prediction_performance}(\textbf{A}), the one-day-ahead predictions obtained from the LSTM model are compared with the actual pressure head values in the test dataset. From this figure, it is evident that the identified LSTM model is able to accurately model the  root zone pressure head in the loamy sand soil column while capturing its general trend. In model predictive algorithms, it is required that at any given time, the process outputs be predicted many time-steps into the future. To this end, the identified LSTM model is used to predict the root zone pressure head for long periods of time. Particularly, these multistep-ahead predictions are produced recursively by iterating the  identified one-step-ahead LSTM model in which previously predicted pressure head values are used as inputs in successive predictions. This is done for a period of 150 days and the results are depicted in Figure \ref{fig:prediction_performance}(\textbf{B}). It can be seen that the recursive use of the identified LSTM model produces accurate pressure head predictions and the predictive performance is comparable to that of the 1D Richards equation. Notwithstanding the accurate predictions obtained in the multistep-ahead application, the accumulation and propagation of prediction errors in this recursive implementation results in a slight drop in the prediction performance compared to the one-step-ahead prediction. From Figure \ref{fig:prediction_performance}, it can be concluded that the method outlined in section \ref{lstm_model} produces an LSTM model that can be regarded as a good representation of the 1D Richards Equation.

\subsection{Irrigation scheduling in homogeneous fields - Case Study 1}
\subsubsection{Scenario 1}
\begin{figure}[H]\label{fig:open_loop}
	\subfloat[Capillary pressure head.] {
		\begin{minipage}[c][0.8\width]{
				0.45\textwidth}\label{fig:states_ol} 
			\centering
			\includegraphics[width=1\textwidth]{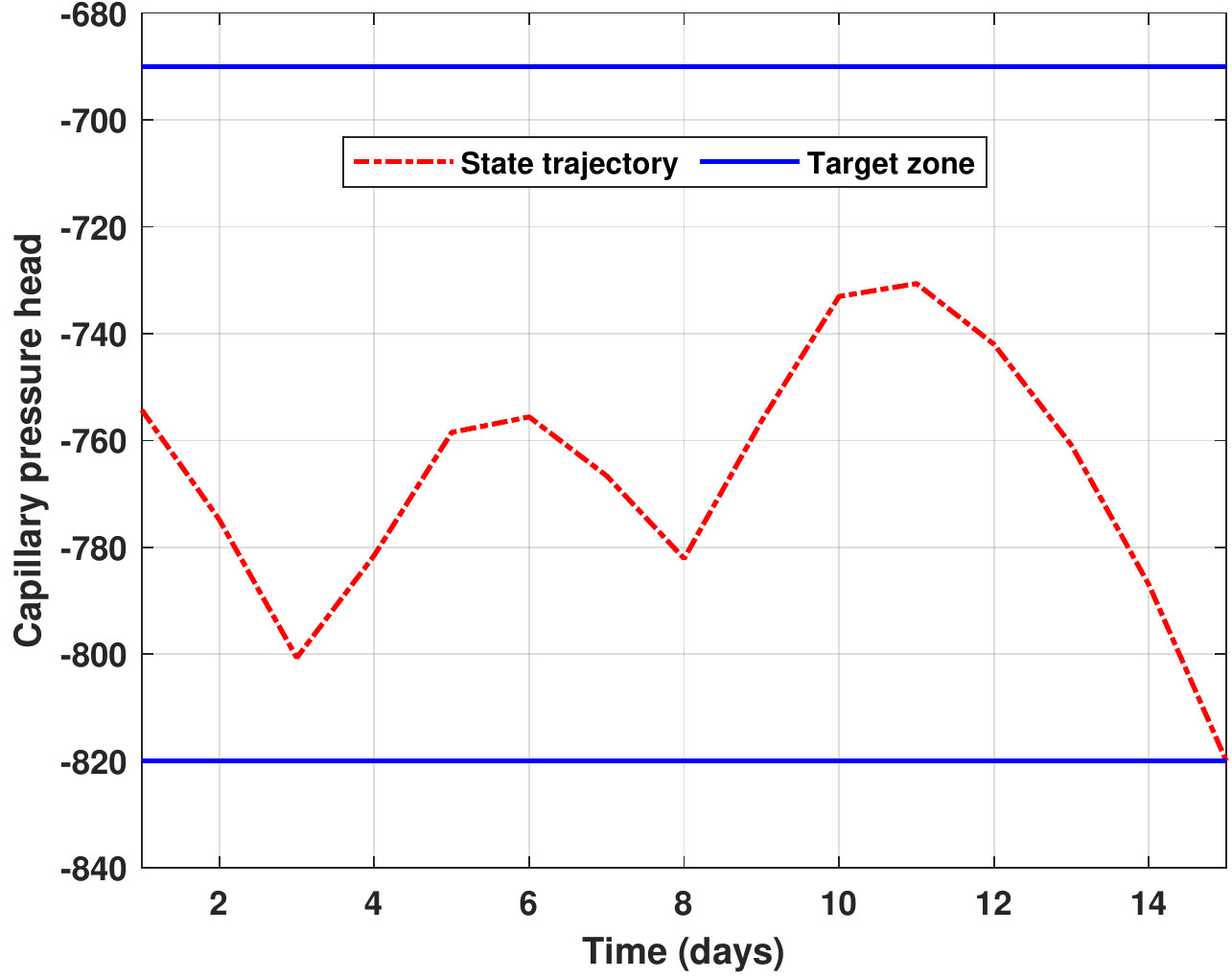}
	\end{minipage}}
	\hfill
	\subfloat[Irrigation amount.]{
		\begin{minipage}[c][0.8\width]{
				0.45\textwidth}\label{fig:control_ol}
			\centering
			\includegraphics[width=1\textwidth]{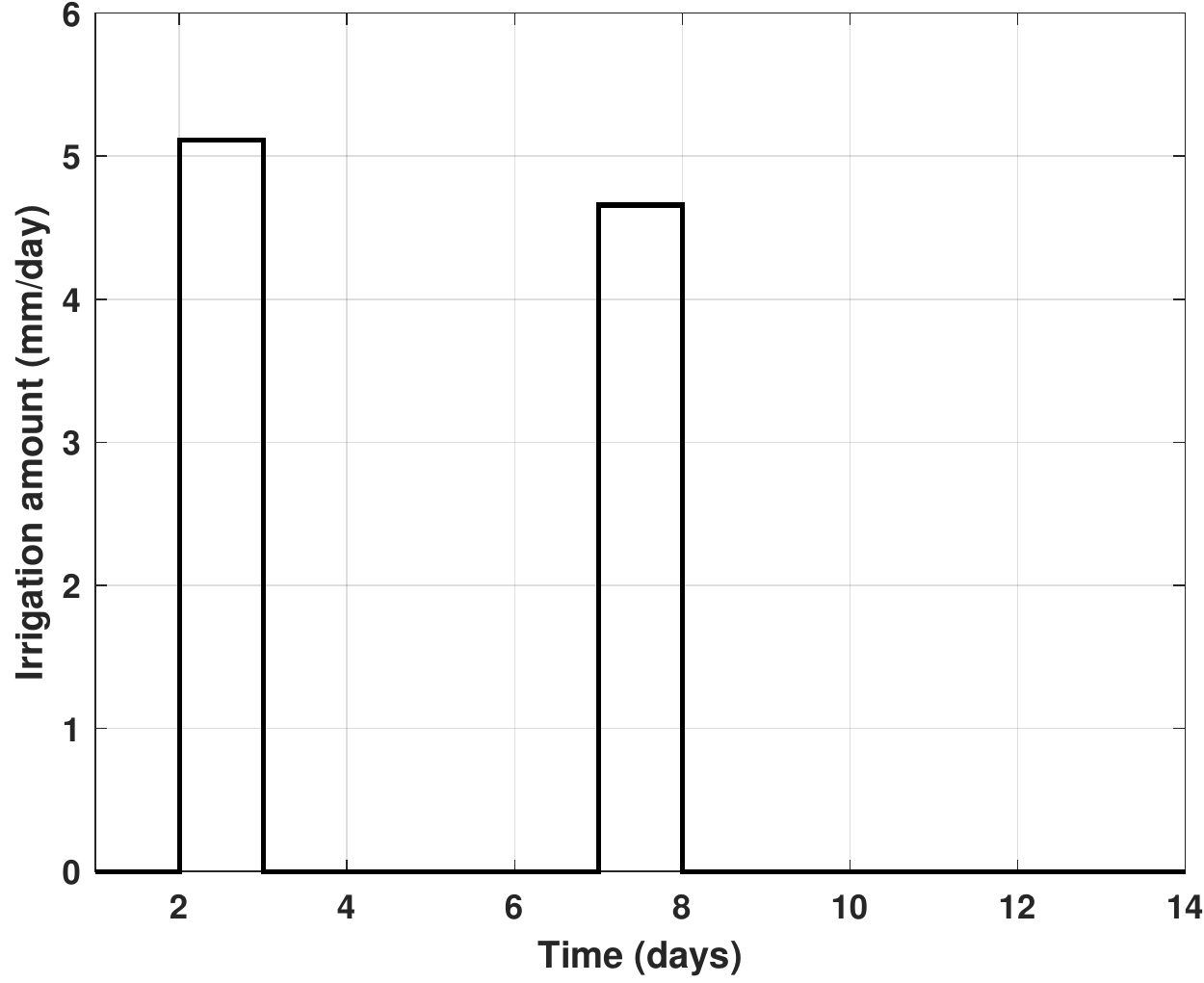}
	\end{minipage}}
	\caption{Open-loop trajectories computed by the scheduler of Equations (\ref{eq:obj}) - (\ref{eq:cons6}) in Scenario 1.}
\end{figure}

\begin{figure}[H]\label{fig:closed_loop}
	\subfloat[Capillary pressure head.] {
		\begin{minipage}[c][0.8\width]{
				0.45\textwidth}\label{fig:states_cl}
			\centering
			\includegraphics[width=1\textwidth]{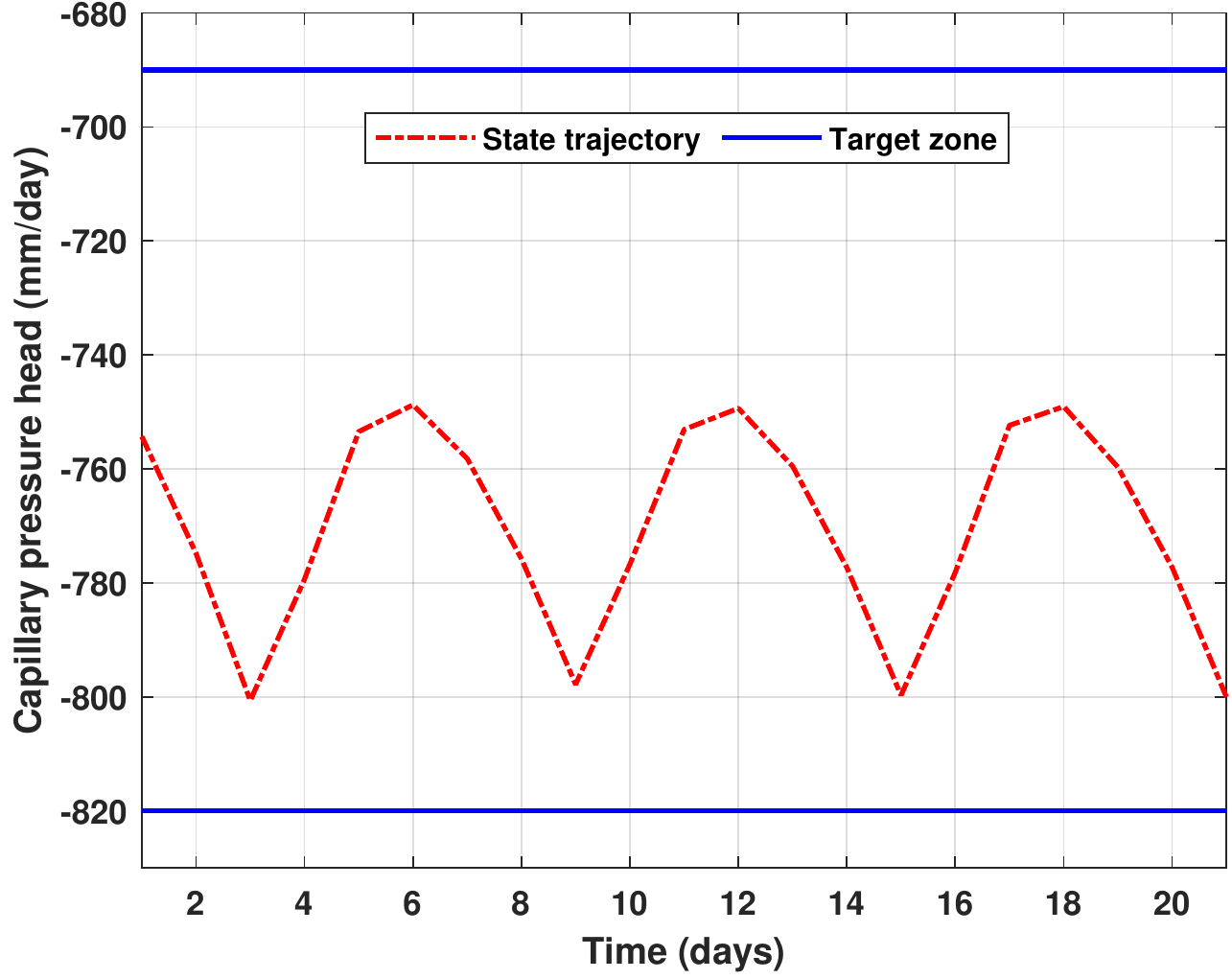}
	\end{minipage}}
	\hfill
	\subfloat[Irrigation amount.]{
		\begin{minipage}[c][0.8\width]{
				0.45\textwidth}\label{fig:contol_cl}
			\centering
			\includegraphics[width=1\textwidth]{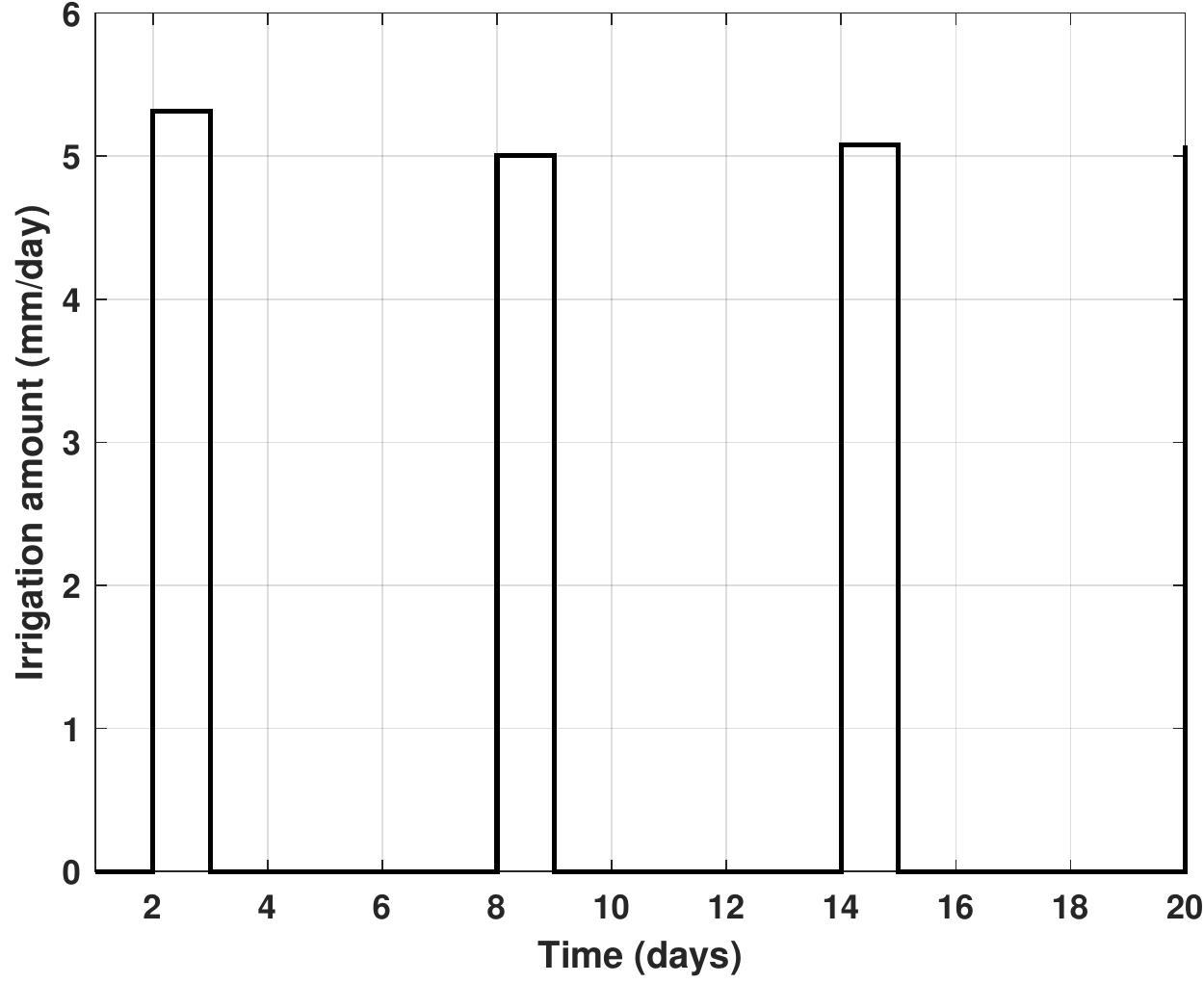}
	\end{minipage}}
	\caption{Closed-loop trajectories under the scheduler of Equations (\ref{eq:obj}) - (\ref{eq:cons6}) in Scenario 1. }
\end{figure}

 Firstly, results of a single evaluation (open-loop simulation) of the scheduler are analyzed. Figure \ref{fig:states_ol} shows the predicted state trajectory under the optimal input trajectory in Figure \ref{fig:control_ol}. It can be seen that by prescribing irrigation amounts of 5.2 mm/day and  4.8 mm/day on days 3 and 7, respectively, the irrigation scheduler is able to maintain the root zone pressure head in the target zone. Applying the open-loop predicted input sequence to the field will usually result in a deviation between the predicted root zone pressure head and the actual dynamics of the root zone pressure head. This deviation will be due to a possible mismatch between the actual field and the identified LSTM model, and additive disturbances arising from errors in weather forecast data. To this end, we consider a closed-loop implementation of the scheduler known as the receding horizon control (RHC) scheme which is known to provide some degree of inherent robustness to these uncertainties. In this framework, only the first control input of the optimal control sequence is implemented, and, to incorporate feedback into the control strategy, the process is repeated at the next time instant using newly obtained information of the state. In the RHC strategy of this case study, the scheduler is evaluated each day for a period of 20 days. Figures \ref{fig:states_cl} and \ref{fig:contol_cl} show the trajectories obtained when the scheduler is operated in closed-loop. In the closed-loop operation of the scheduler, it prescribes irrigation amounts of 5.3 mm/day, 5.0 mm/day, 5.1 mm/day, and 5.0 mm/day on days 3, 8, 15, and 20,  to maintain the root zone pressure head in the target zone.

\subsubsection{Scenario 2}
\begin{figure}[H]
	\subfloat[Capillary pressure head.] {
		\begin{minipage}[c][0.8\width]{
				0.45\textwidth}\label{fig:states_ol_rain} 
			\centering
			\includegraphics[width=1\textwidth]{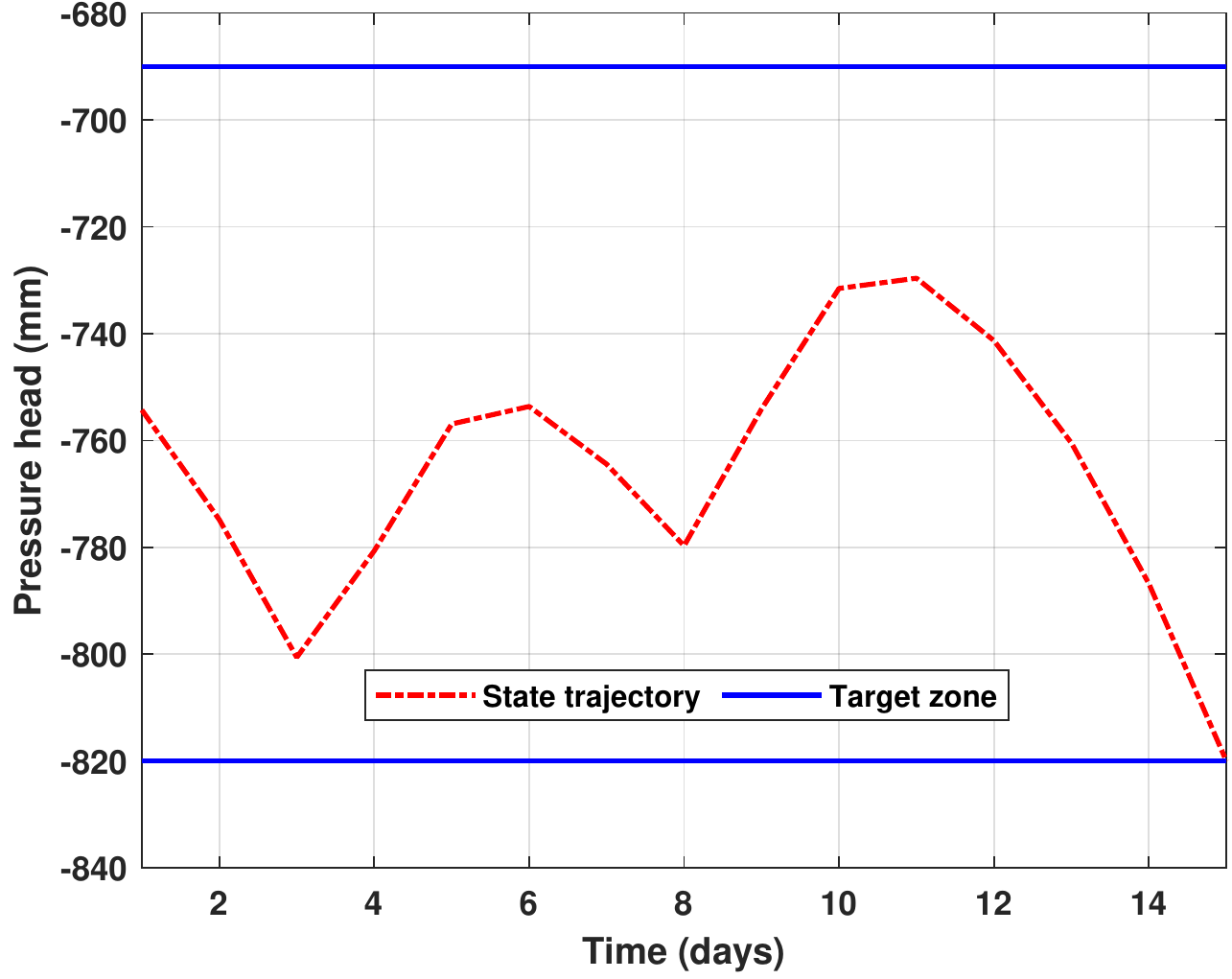}
	\end{minipage}}
	\hfill
	\subfloat[Irrigation amount.]{
		\begin{minipage}[c][0.8\width]{
				0.45\textwidth}\label{fig:control_ol_rain}
			\centering
			\includegraphics[width=1\textwidth]{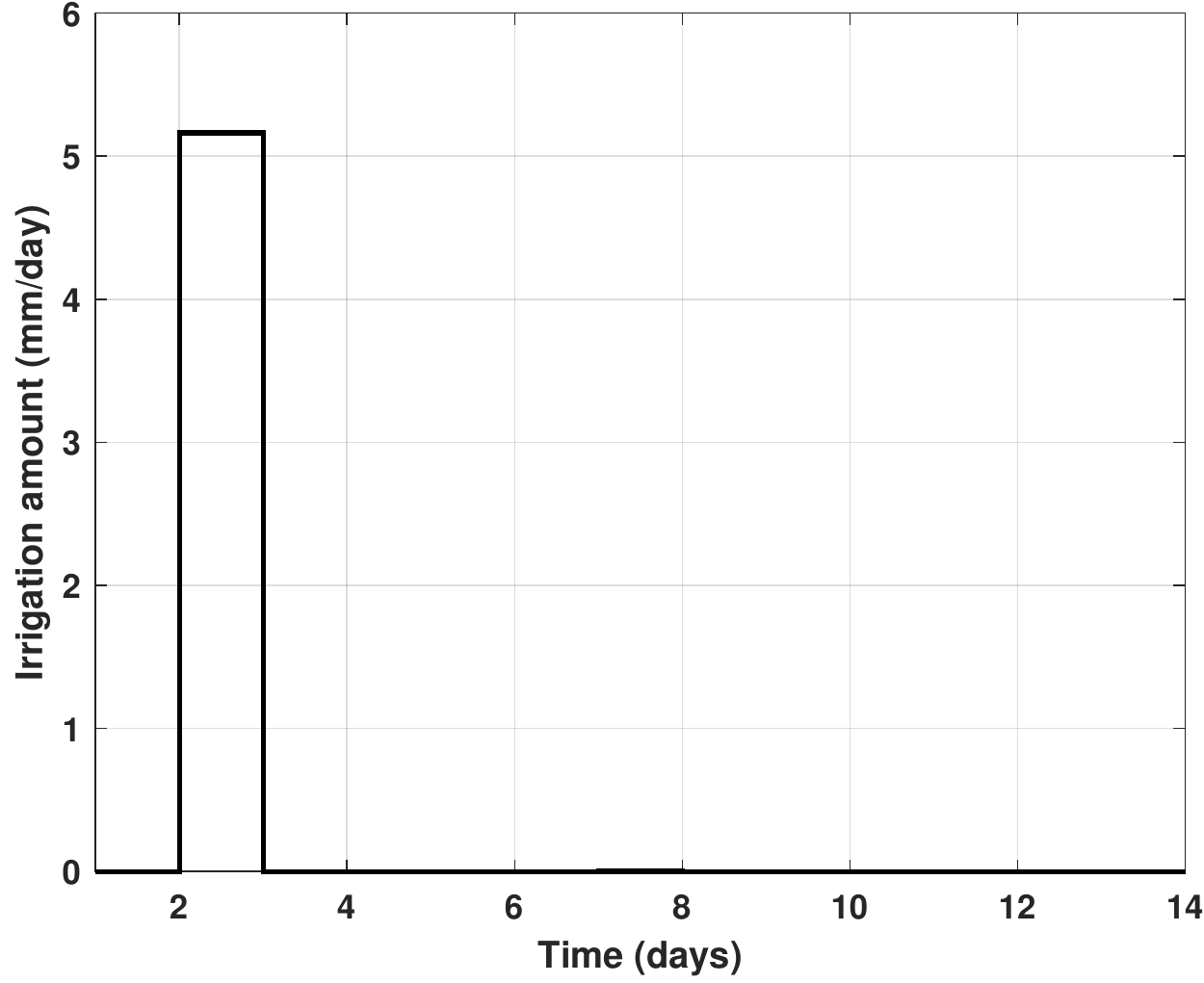}
	\end{minipage}}
	\caption{Open-loop trajectories computed by the scheduler of Equations (\ref{eq:obj}) - (\ref{eq:cons6}) in Scenario 2.}
\end{figure}

\begin{figure}[H]
	\subfloat[Capillary pressure head.] {
		\begin{minipage}[c][0.8\width]{
				0.45\textwidth}\label{fig:states_cl_rain}
			\centering
			\includegraphics[width=1\textwidth]{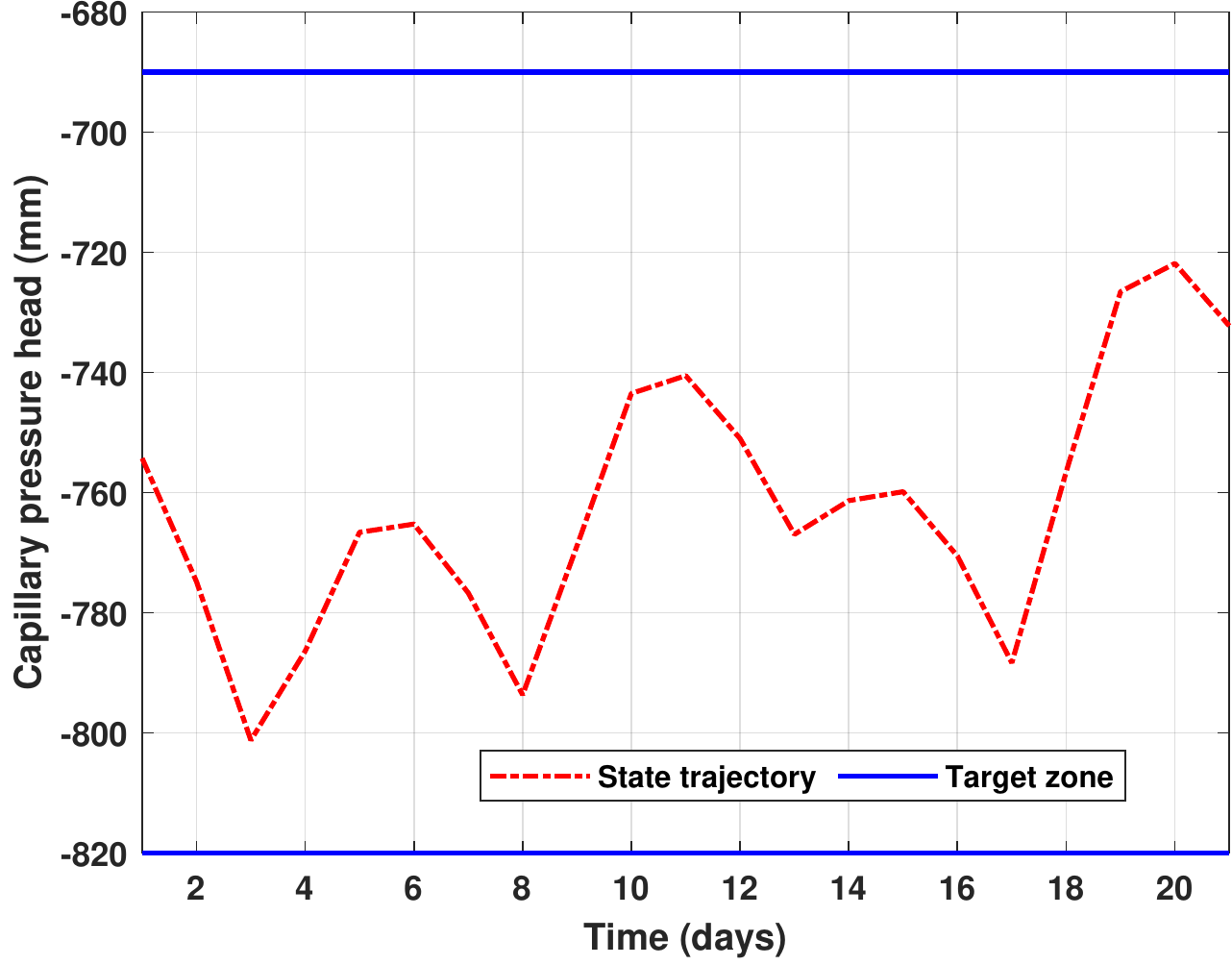}
	\end{minipage}}
	\hfill
	\subfloat[Irrigation amount.]{
		\begin{minipage}[c][0.8\width]{
				0.45\textwidth}\label{fig:control_cl_rain}
			\centering
			\includegraphics[width=1\textwidth]{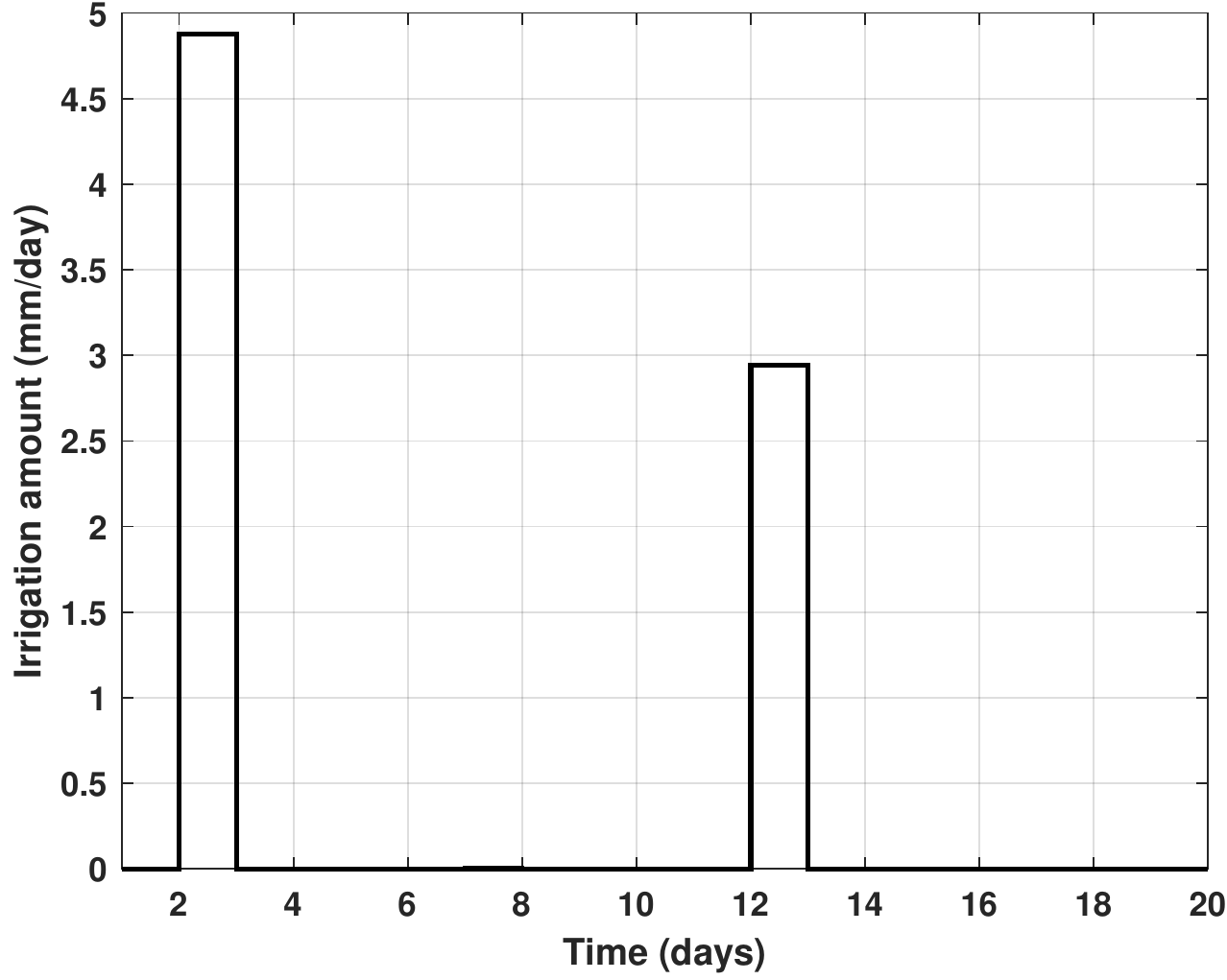}
	\end{minipage}}
	\caption{Closed-loop trajectories under the scheduler of Equations(\ref{eq:obj}) - (\ref{eq:cons6}) in Scenario 2. }
\end{figure}

In this scenario, precipitation values of 4.8 mm and 5.2 mm are considered on days 6 and 16 of the simulation period. It can be seen from the open-loop (Figures \ref{fig:states_ol_rain} \& \ref{fig:control_ol_rain}) and closed-loop (Figures \ref{fig:states_cl_rain} \& \ref{fig:control_cl_rain}) implementations of the scheduler that in the presence of rain, the scheduler prescribes less frequent irrigation, compared to the corresponding implementations in Scenario 1. As a result of the anticipated rain on the 6th day, the open-loop implementation of the scheduler prescribes an irrigation amount of 5.2 mm/day on the 6th day (Figure \ref{fig:control_ol_rain}) and this together with the rain forecast is able to keep the root zone pressure head in the target zone (Figure \ref{fig:states_ol_rain}). Similarly, due to the anticipated rain on days 6 and 16, the closed-loop implementation of the scheduler prescribes 4.9 mm/day and 2.9 mm/day of irrigation on days 3 and 13 (Figure \ref{fig:control_cl_rain}) to maintain the root zone pressure head in the target zone (Figure \ref{fig:states_cl_rain}).

\subsection{Approximating the binary variable with a sigmoid function} 
\begin{figure}[H]
	\subfloat[Capillary pressure head.] {
		\begin{minipage}[c][0.80\width]{
				0.450\textwidth}\label{fig:states_ol_sig}
			\centering
			\includegraphics[width=1\textwidth]{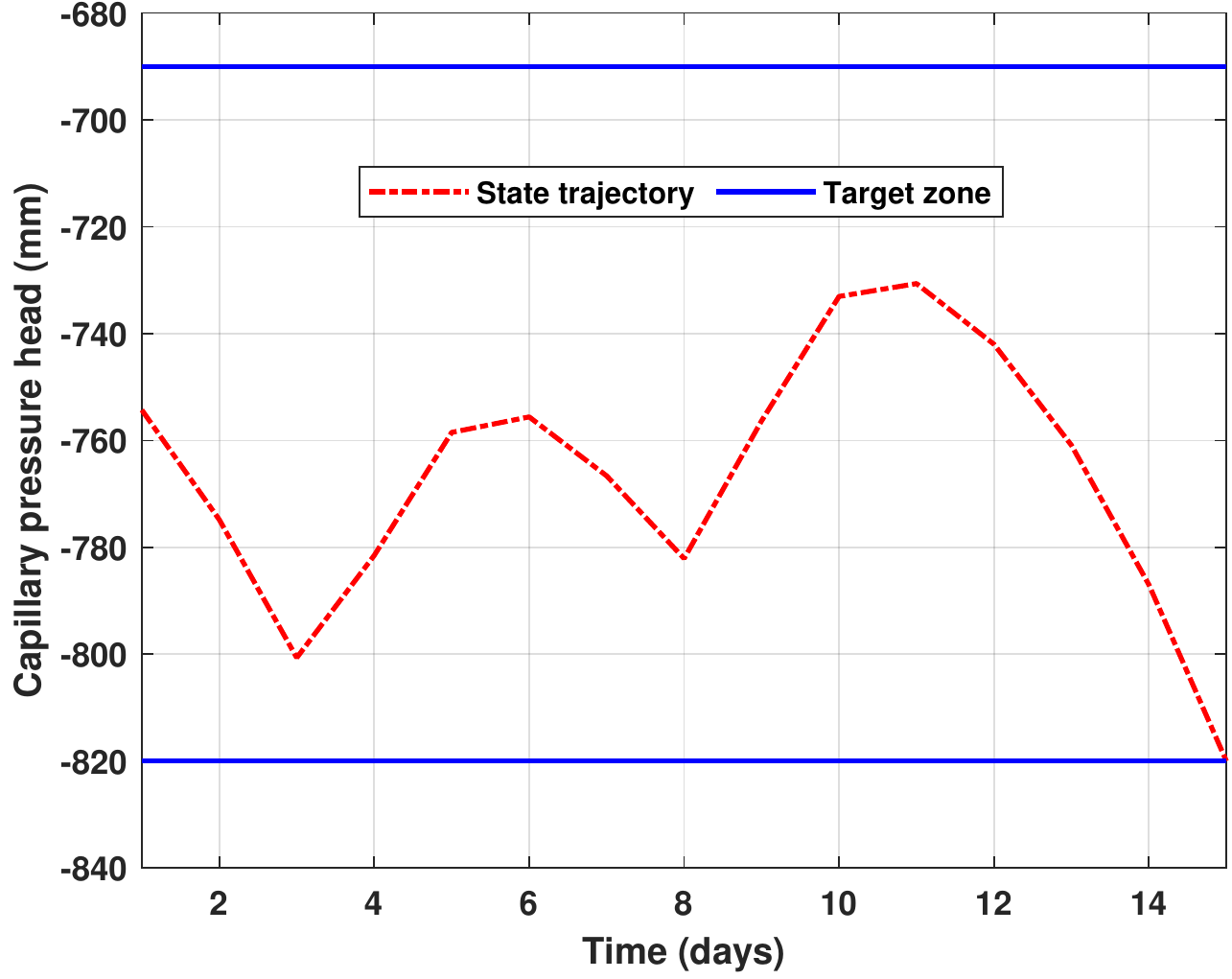}
	\end{minipage}}
	\hfill
	\subfloat[Irrigation amount.]{
		\begin{minipage}[c][0.80\width]{
				0.450\textwidth}\label{fig:control_ol_sig}
			\centering
			\includegraphics[width=1\textwidth]{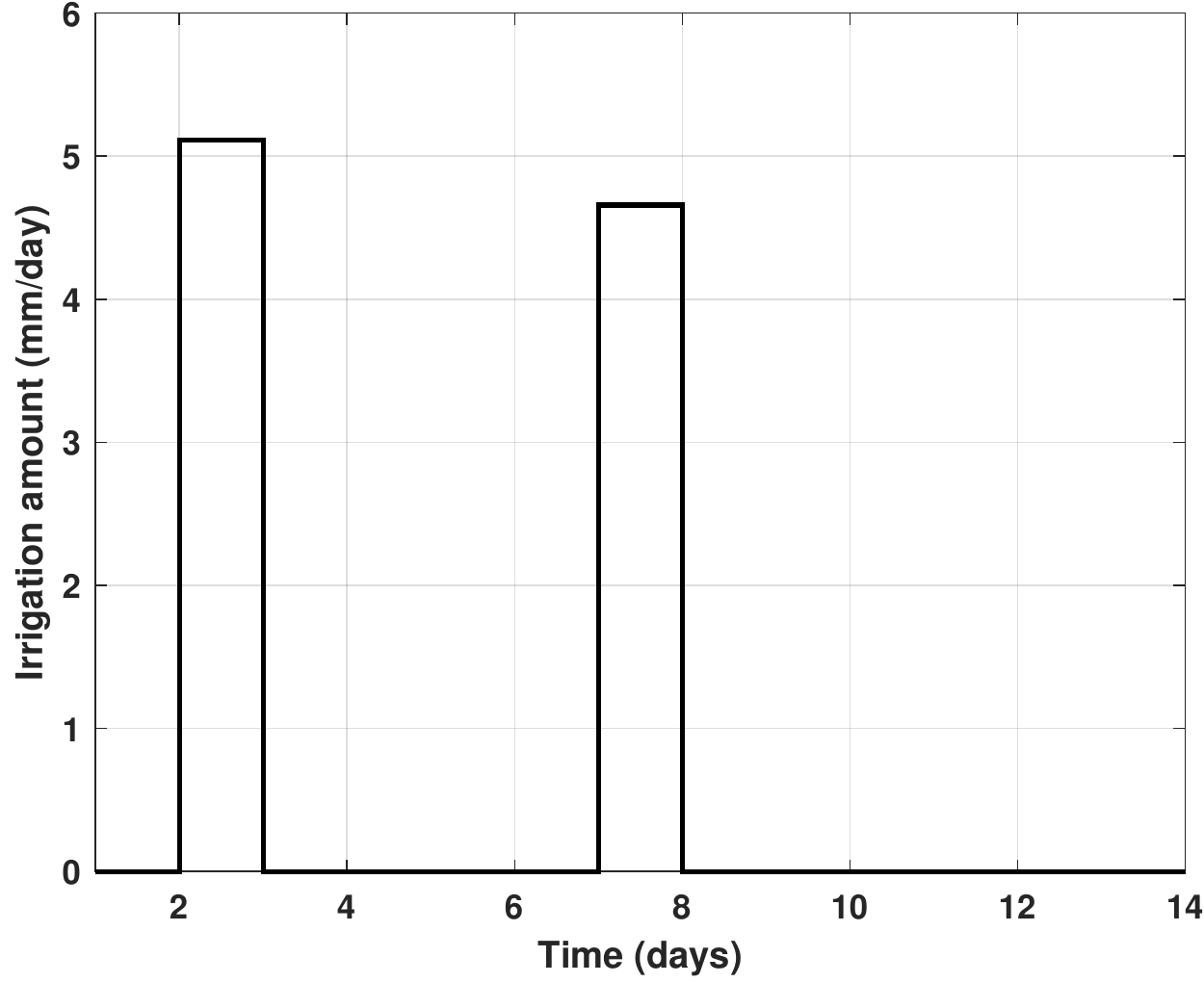}
	\end{minipage}}
	\caption{Open-loop trajectories computed by the scheduler of Equations (\ref{eq:obj_sig}) - (\ref{eq:cons6_sig}).}
\end{figure}

\begin{figure}[H]
	\subfloat[Capillary pressure head.] {
		\begin{minipage}[c][0.80\width]{
				0.45\textwidth}\label{fig:contol_cl_sig}
			\centering
			\includegraphics[width=1\textwidth]{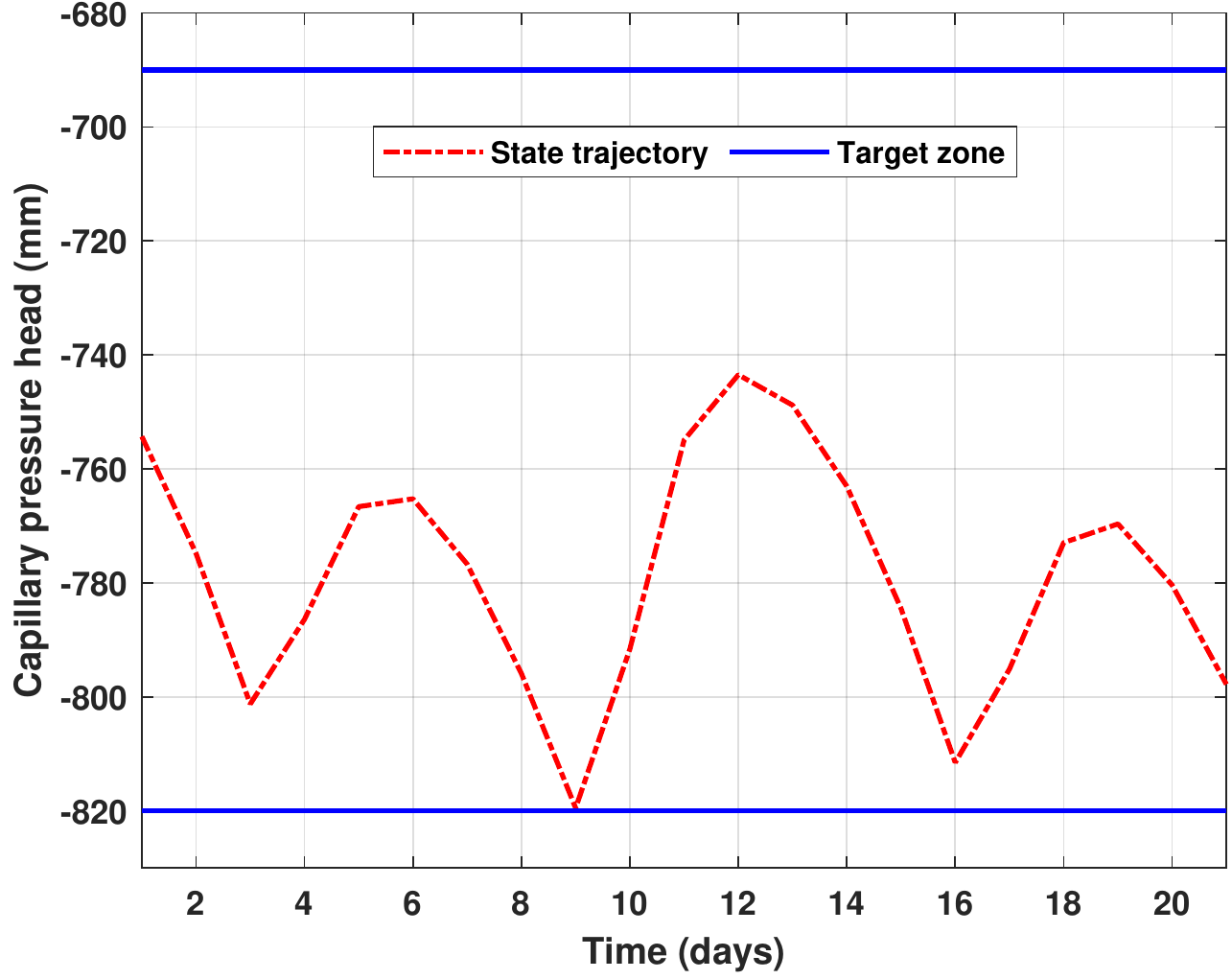}
	\end{minipage}}
	\hfill
	\subfloat[Irrigation amount.]{
		\begin{minipage}[c][0.80\width]{
				0.45\textwidth}\label{fig:state_cl_sig}
			\centering
			\includegraphics[width=1\textwidth]{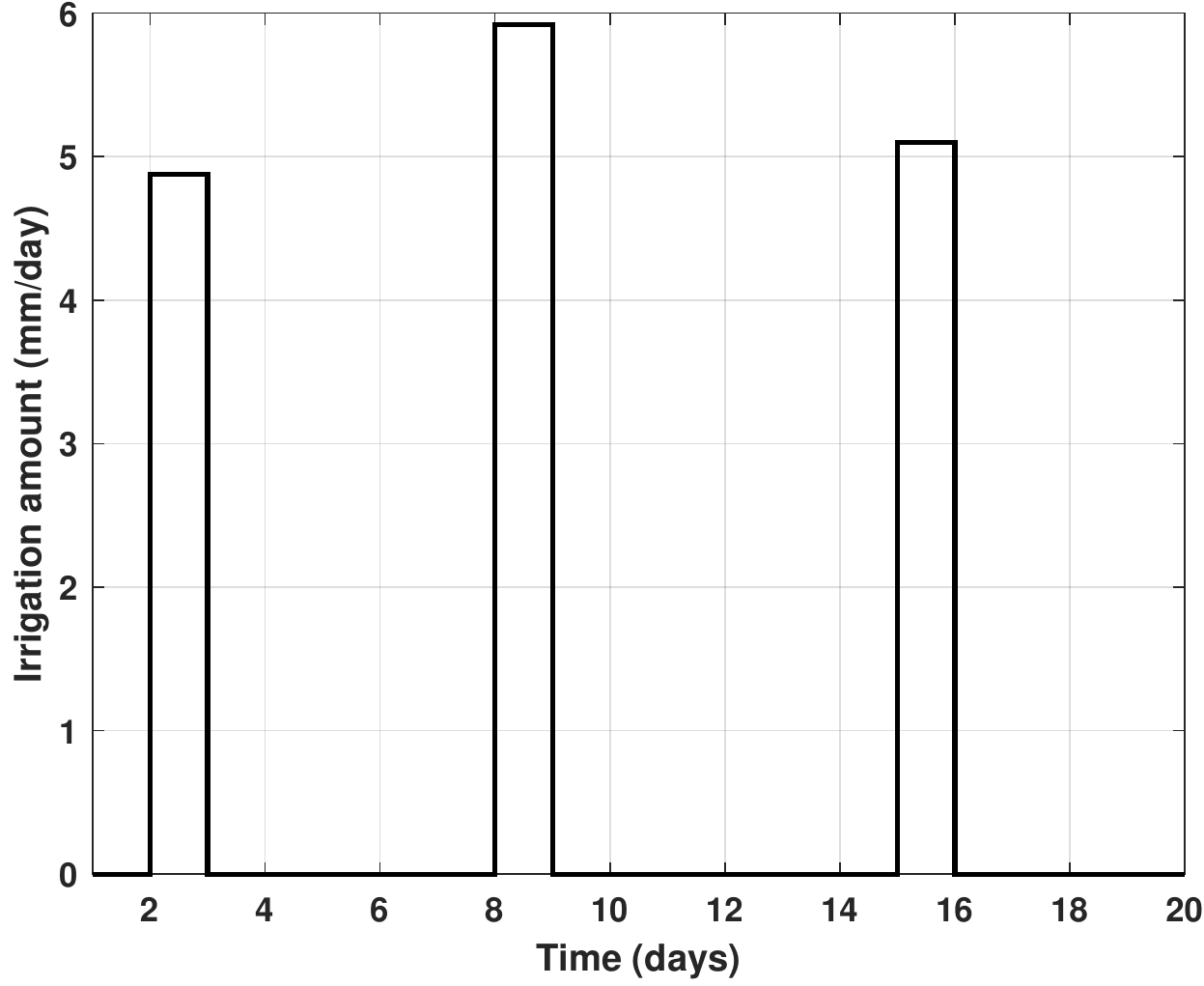}
	\end{minipage}}
	\caption{Closed-loop trajectories under the scheduler of Equations (\ref{eq:obj_sig}) - (\ref{eq:cons6_sig}). }
\end{figure}

In this section, the performance of the sigmoid approximation is investigated under Scenario 1 of Case Study 1. This simulation study revealed that a slope of 25 was able to provide a good convergence of the sigmoid function to the binary elements while preventing ill-conditioning issues. A visual inspection of Figures \ref{fig:control_ol_sig} and \ref{fig:states_ol_sig} reveals that, in the open-loop setting, the sigmoid function approximation gives similar results as the mixed-integer formulation (Figures \ref{fig:control_ol} and \ref{fig:states_ol}). Table \ref{tbl:comparison_ol} further reveals that the numerical values of the cost function in both formulations are identical. In the closed-loop implementation, the results summarized in Figures \ref{fig:contol_cl_sig} \& \ref{fig:state_cl_sig} demonstrate that the results obtained from the sigmoid function approximation are comparable to that which were obtained in the mixed-integer formulation (Figures \ref{fig:contol_cl} \& \ref{fig:states_cl}). The numerical values of the cost in Table \ref{tbl:comparison_cl} reveal that with a careful selection of the sigmoid function's slope according to Algorithm \ref{alg:cap}, the formulation involving the sigmoid function can produce a reduced cost compared to the mixed-integer formulation. 
With respect to the computation time, it is evident from Tables \ref{tbl:comparison_ol} \& \ref{tbl:comparison_cl} that the computation speed of the proposed scheduler can be remarkably enhanced when the binary variable in the original formulation is approximated with a sigmoid function.

\begin{table}[t]
	\caption{Comparison between the mixed-integer and sigmoid formulations (Open-loop).}
	\small 
	\centering
	\begin{tabular}{ccc}
		\hline
		\textbf{Formulation}& Mixed-Integer & Sigmoid Approximation \\
		\hline
		\textbf{Computation Time (minutes)}& 57.8 & 2.5 \\
		\hline
		\textbf{Cost}&106.5&106.5\\
		\hline
	\end{tabular} \label{tbl:comparison_ol}
\end{table}

\begin{table}[t]
	\caption{Comparison between the mixed-integer and sigmoid formulations (Closed-loop).}
	\small 
	\centering
	\begin{tabular}{ccc}
		\hline
		\textbf{Formulation}& Mixed-Integer & Sigmoid Approximation \\
		\hline
		\textbf{Computation Time (hours)}& 10.5 & 2.0 \\
		\hline
		\textbf{Cost}&213.4 &176.6 \\
		\hline
	\end{tabular} \label{tbl:comparison_cl}
\end{table}

\subsection{Spatially variable irrigation scheduling in small-scale fields - Case Study 2}
\begin{figure}[H]
	\centering
	\includegraphics[width=0.9\textwidth]{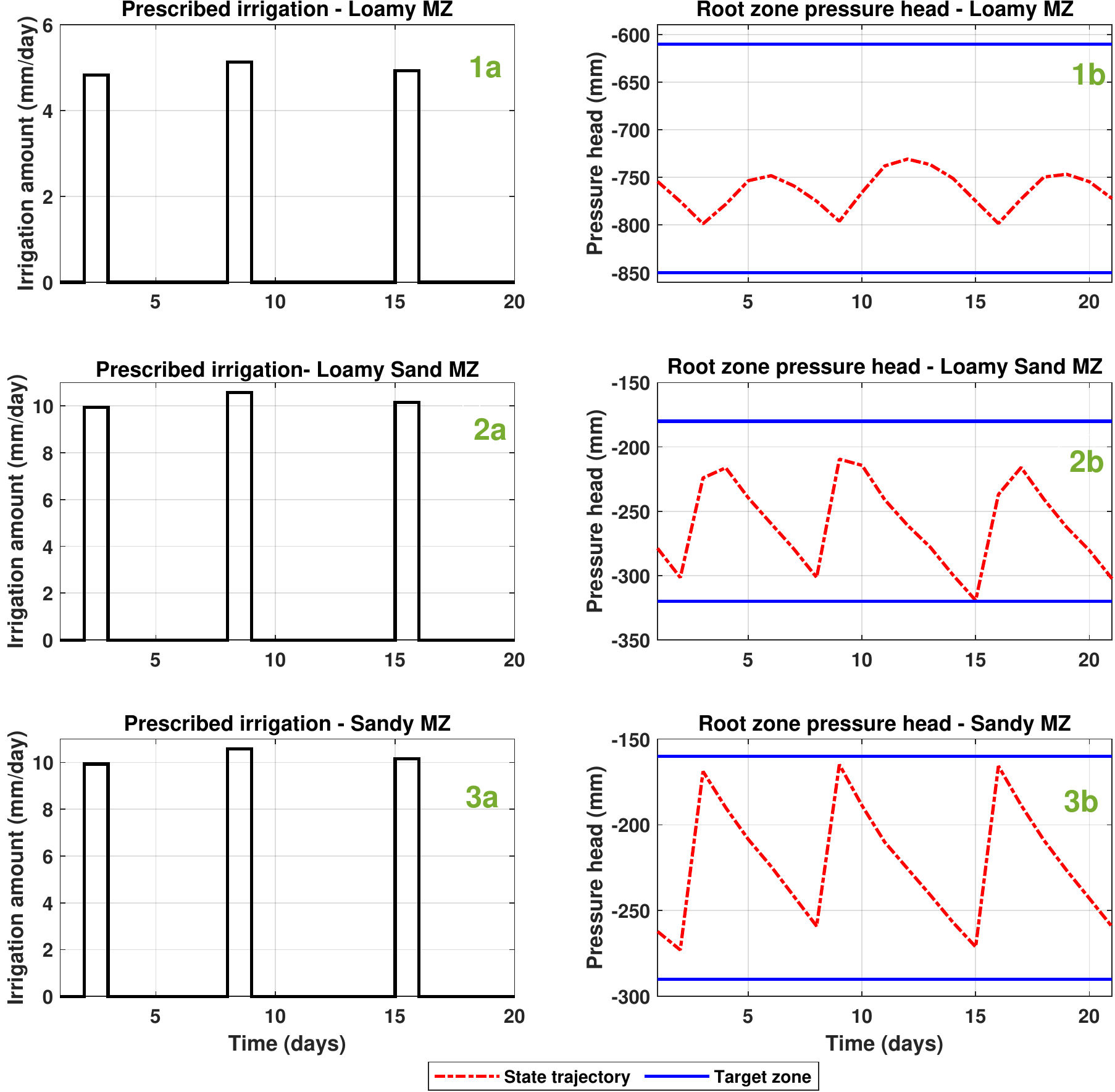}
	\caption{Closed-loop trajectories under the spatially variable scheduler of Equations (\ref{eq:obj_sv}) - (\ref{eq:lastconst_sv}) for (i) loam MZ (\textbf{1a} \& \textbf{1b}), (ii) loamy sand  MZ  (\textbf{2a} \& \textbf{2b}), and (iii) sand MZ (\textbf{3a} \& \textbf{3b}). }
	\label{fig:multi_small} 
\end{figure}

Figure \ref{fig:multi_small} shows the closed-loop scheduling results for a small-scale spatially variable field that is delineated into three MZs.
It is evident that the scheduler prescribes irrigation on days 2, 9, and 16 for all the three MZs. This is due to the fact that the spatially variable scheduler for small-scale fields assumes that the irrigation implementing equipment is able to irrigate all the MZs in one day. Thus, whenever the discrete irrigation decision is equal to 1, all the three MZs must be irrigated. In the loam MZ, the scheduler prescribes irrigation amounts of 4.8 mm/day, 5.2 mm/day, and 4.9 mm/day to maintain the root zone pressure head within the target zone. In the loamy sand and sandy soil MZs, the scheduler prescribes irrigation amounts of 10 mm/day, 10.5 mm/day and 10.1 mm/day to sustain the root zone pressure head in the target zone. The scheduler prescribed same irrigation amounts for these MZs due to the fact that loamy sand soils and sandy soils posses similar infiltration and drainage features. 

\subsection{Spatially variable irrigation scheduling in large-scale fields - Case Study 3}
\begin{figure}[H]
	\centering
	\includegraphics[width=0.8\textwidth]{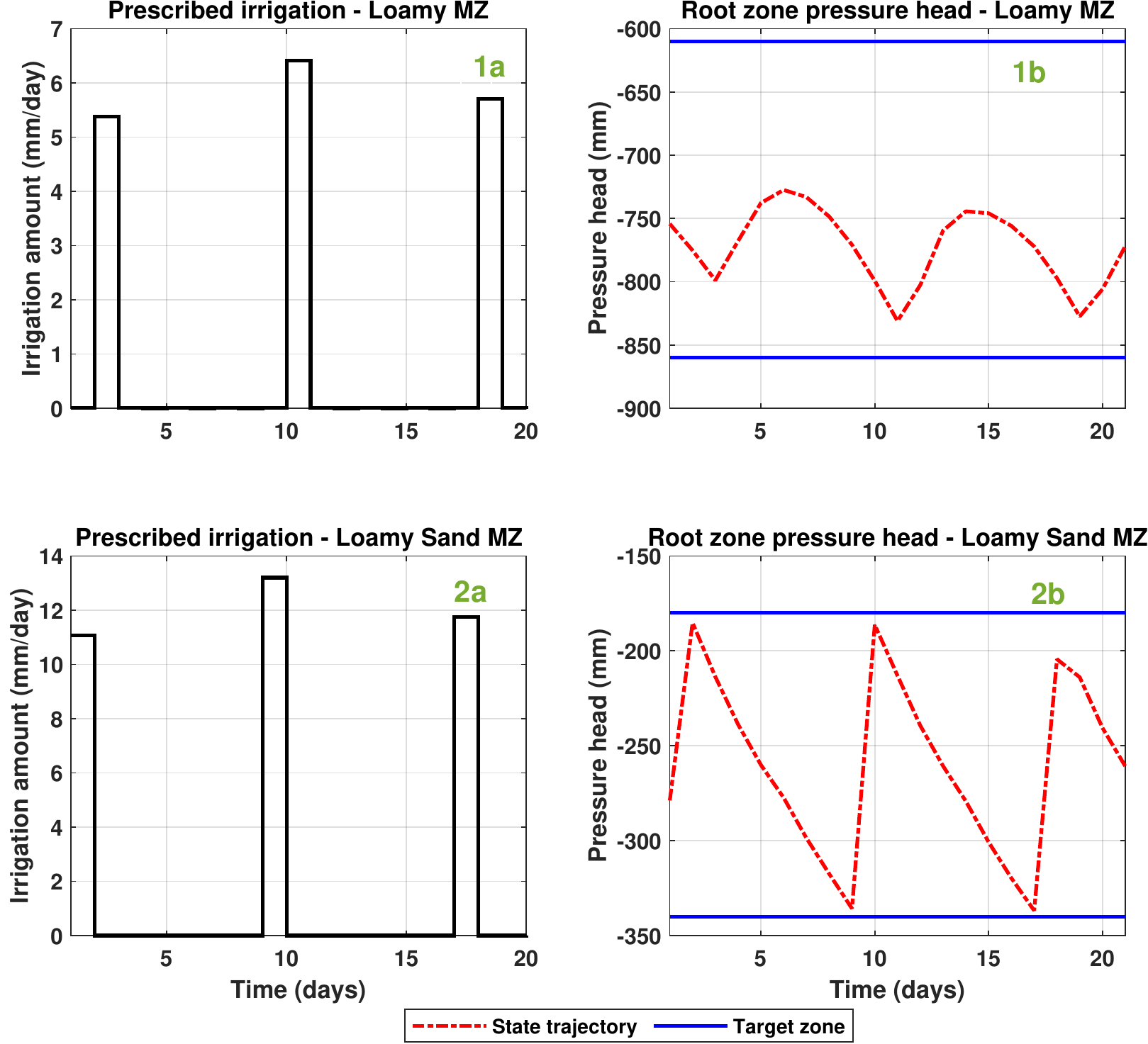}
	\caption{Closed-loop trajectories under the spatially variable scheduler of Equations (\ref{eq:obj_sv_bg}) - (\ref{eq:lastconst_sv_bg}) for (i) loam MZ (\textbf{1a} \& \textbf{1b}), and (ii) loamy sand MZ  (\textbf{2a} \& \textbf{2b}). }
	\label{fig:multi_large} 
\end{figure}

Figure \ref{fig:multi_large} shows the closed-loop scheduling results for a large-scale spatially variable field that is delineated into two MZs. In this simulation case study, the irrigation implementing equipment starts its irrigation cycle from the loamy sand MZ and completes its irrigation cycle after 2 days in the loam MZ. The loam MZ can be irrigated on days 2, 4, 6, 8, 10, 12, 14, 16, 18, and 20 while the loamy sand MZ can be irrigated on days 1, 3, 5, 7, 9, 11, 13, 15, 17, and 19. The scheduler prescribes 3 irrigation cycles during the 20 day simulation period. From Figure \ref{fig:multi_large}, it can be seen that the days that make up the 3 cycles are days 1 and 2 for the first cycle, days 9 and 10 for the second cycle, and days 17 and 18 for the third cycle. For the loam sand MZ, the scheduler prescribes irrigation amounts of 11 mm/day, 13 mm/day, and 12.8 mm/day on days 1, 9, and 17 to maintain the root zone pressure head in the target zone. For the loam MZ, the scheduler prescribes irrigation amounts of 5.4 mm/day, 6.3 mm/day, and 6.2 mm/day on days 2, 10, and 18. 
\section{Summary and conclusions}
In this study, an LSTM-based mixed-integer MPC with zone control for irrigation scheduling was proposed and tested. The proposed scheduling framework seeks to ensure optimal water uptake in crops while minimizing total water consumption and irrigation costs. To this end, an LSTM model was developed to describe the dynamics of the root zone capillary pressure head. This data-driven machine learning model was trained using open-loop simulated data from the Richards equation. A mixed-integer MPC with zone control was then developed using the identified LSTM model. A heuristic method using the sigmoid function was proposed to simplify the mixed-integer MPC in order to reduce the evaluation time of the scheduler. Due to the prevalence of spatial variability in many agricultural fields, this study went ahead to adapt the proposed scheduler to small- and large-scale spatially variable fields. 

The simulation results revealed that, the LSTM model was capable of performing accurate single-step and multi-step predictions of the root zone capillary pressure head. The results from the simulation case studies highlight the efficacy of the proposed scheduler, as it was able to prescribe irrigation schedules that are typical of irrigation practice on most irrigated farms. The proposed approach can thus be successfully used to maximize crop yield while minimizing the total water consumption and irrigation costs. The heuristic method involving the sigmoid function was capable of enhancing the computational efficiency of the scheduler and this underscores the capability of the proposed approach to prescribe optimal or near-optimal irrigation schedules within workable computational budgets.
\section*{Acknowledgment}
Financial support from Alberta Innovates and Natural Sciences and Engineering Research Council of Canada is gratefully acknowledged.  

\end{document}